\documentclass[useAMS,usenatbib,usegraphicx,useepstopdf]{mn2e} % for pdflatex
\usepackage{hhline}
\usepackage{color}
\usepackage{ulem}
\usepackage{epstopdf}  % for pdflatex

\def\mbf#1{\mbox{\boldmath ${#1}$}}

\def\lesssim{\; \buildrel < \over \sim \;}
\def\gtrsim{\; \buildrel > \over \sim \;}

\newcommand\aj{{AJ}}% 
\newcommand\araa{{ARA\&A}}% 
\newcommand\apj{{ApJ}}% 
\newcommand\apjl{{ApJ}}% 
\newcommand\apjs{{ApJS}}% 
\newcommand\aap{{A\&A}}% 
\newcommand\mnras{{MNRAS}}% 
\newcommand\pasj{{PASJ}}% 
\newcommand\nat{{Nature}}% 
\newsavebox\myVerb

\title[Noncircular Motion and Outflows in the Galactic Bulge]{Stochastic Noncircular Motion and Outflows Driven 
\\by Magnetic Activity in the Galactic Bulge Region}
\author[Suzuki et al.]{Takeru K. Suzuki$^{1}$\thanks{E-mail:
stakeru@nagoya-u.jp (TKS)},Yasuo Fukui$^{1}$,Kazufumi Torii$^{1}$, 
Mami Machida$^{2}$,\& Ryoji Matsumoto$^{3}$\\
$^{1}$Department of Physics, Nagoya University,Furo-cho, Chikusa-ku, Nagoya, Aichi 464-8602, Japan\\
$^{2}$Department of Physics, Faculty of Sciences, Kyushu University,
6-10-1 Hakozaki, Higashi-ku, Fukuoka 812-8581, Japan \\
$^{3}$Department of Physics, Graduate School of Science, Chiba University,
1-33 Yayoi-cho, Inage-ku, Chiba 263-8522, Japan}

\begin{document}
%\begin{CJK*}{UTF8}{gbsn}

\pagerange{\pageref{firstpage}--\pageref{lastpage}} \pubyear{2015}

\maketitle

\label{firstpage}

\begin{abstract}
%It is often discussed that observed noncircular motion of the gas in the 
%Galactic center region is a consequence of a stellar bar-like potential.  
%We propose an alternative scenario involving magneto-hydrodynamical turbulence.
By performing a global magneto-hydrodynamical simulation for the Milky Way 
with an axisymmetric gravitational potential, 
we propose that spatially dependent amplification of magnetic fields possibly 
explains the observed noncircular motion of the gas in the Galactic center 
region.
%A region with relatively small rotation frequency is formed in $0.3-1$ kpc 
%to satisfy the global equilibrium including the gas pressure. 
The radial distribution of the rotation frequency in the bulge region is  
not monotonic in general.  
The amplification of the magnetic field is enhanced in regions with stronger 
differential rotation, because magnetorotational instability and field-line 
stretching are more effective. 
%As a result, the amplification of the magnetic field occurs in an inhomogeneous 
%manner, %bumpy structure in the radial direction, 
%An inevitable consequence is 
%the steep decline of the magnetic energy with increasing 
%Galactocentric distance, and the associated 
%and the associated magnetic pressure-gradient force drives radial flows 
%in a stochastic manner.  
The strength of the amplified magnetic field reaches $\gtrsim 0.5$ mG, and 
radial flows of the gas are excited by the inhomogeneous transport 
of angular momentum through turbulent magnetic field that is amplified in a 
spatially dependent manner. 
In addition, the magnetic pressure-gradient force also drives radial flows 
in a similar manner. 
As a result, the simulated position-velocity diagram exhibits a time-dependent 
asymmetric parallelogram-shape owing to the intermittency of the magnetic 
turbulence; the present model provides a viable alternative to 
the bar-potential-driven model for the parallelogram-shape of 
the central molecular zone. This is a natural extension 
into the central few 100 pc of the magnetic activity, which 
is observed as molecular loops at radii from a few 100 pc to 1 kpc. 
Furthermore, the time-averaged net gas flow is directed outward, 
whereas the flows are highly time-dependent, which we discuss from a viewpoint 
of the outflow from the bulge. 
%While the toroidal magnetic field dominates in the disk by the radial 
%differential rotation, both the poloidal and toroidal fields are comparably 
%generated in the bulge because of the weak differential rotation. 
%In the central region $< 200$ pc, nonaxisymmetric dense regions with 
%mass $>10^7 M_{\odot}$ and hydrogen number density $> 2000$ cm$^{-3}$ 
%transiently forms, which we 
%discuss as a possible candidate of the central molecular zone.

\end{abstract}

\section{Introduction}
Observations of the atomic and molecular interstellar medium in the 
inner Milky Way exhibit large noncircular motion of the gas 
\citep{ro60,sco72,lb78,bal87,oka98,tsu99,saw04,oka05,tak10}. %,mt00}
Among various possible explanations for the noncircular motion, \citet{dev64} 
raised a possibility that the gas responds to a stellar bar potential 
of the Milky Way. 
\citet{pet75} adopted a model that takes into accout gas flowing along elliptic 
streamlines caused by a bar-like potential and showed that the general trend 
of the neutral hydrogen (HI) distribution is explained by the model. 
\citet{lb80} developed a tilted bar model that accounts for the observed 
HI distributions at different latitudes.

Observationally, \citet{bl91} identified a bar-like distribution 
of stars in the near-infrared distribution
by \citet[][see also \cite{hay81}]{mat82}\footnote{Infrared 
observation of the Galactic center region was also independently done 
by \citet{oku77}.}.
Based on the above consideration and finding, \citet{bin91} showed that a 
bar-like potential naturally reproduces 
the noncircular motion by comparing observed and calculated $l-v$ 
(Galactic longitude -- line-of-sight velocity from the Local Standard of Rest; 
LSR, hereafter) diagrams. 
%However, it is not easy to explain the observed asymmetry 
%of the parallelogram-shape in the $l-v$ diagrams by a simple bar potential.
The streaming motion driven by the bar potential provides 
an only viable explanation on the parallelogram, which has been explicitly 
discussed in literatures \citep{bli94,kw02,rod08,bab10,mol11}. 

We however find  that the observed parallelogram 
in the 2.6mm CO emission shows asymmetry in the positive and negative 
velocities and in $l$, including CO features only 
at positive longitudes, $l=$3 degrees (clump 2) and 5.5 degrees  
\citep[e.g.,][]{ban77,tor10b}. We also see that the central molecular zone 
(CMZ, hereafter) shows vertical CO features up to vertical distance 
$\sim 200$ pc \citep{eno14,tor14a,tor14b}, which we discuss further later 
in this section. We note that these outstanding observed 
characteristics remain unexplored by the bar potential model.
%Since then, roles of various types of bar potentials in the bulge have been 
%extensively studied to date \citep{kw02,rod08,bab10,mol11}.

We note that few efforts have actually been made to elaborate 
detailed observed features of the parallelogram until recently by numerical 
simulations in the bar potential. \citet{rod08} presented the first numerical 
simulations for the parallelogram. Their Figure 14 shows 
a simulated parallelogram to be compared to the CMZ, while 
their result is yet crude at best with no detailed fitting to the gas 
distribution in the $l$-$v$ diagram. These authors concluded that the 
observed asymmetry of the CMZ cannot be due to lopsidedness of the bar 
and suggested that the asymmetry may be of an ad-hoc origin like infalling 
gas into the CMZ at $l$=1.3$^\circ$. Their conclusion indicates 
that the bar potential alone cannot reproduce the asymmetry of the CMZ 
or the high-z CO features including the 1.3$^\circ$ complex. The difficulty 
in explaining the CMZ makes a contrast with the large-scale simulations 
in the bar potential over several kpc, which successfully reproduce the 3-kpc 
arm and other major features outside the central kpc \citep[e.g.,][]{rod08}.
%(e.g., Rodriguez-Fernandez and Combes 2014).

On the other hand, magnetic field is also supposed to play an important role 
in the Galactic bulge. 
Complicated structures, e.g., nonthermal filaments including Radio Arc, 
are observed in the Galactic center region, which %{\color{blue}are} 
probably attribute to magnetic fields
\citep{yus84,tsu86,chu03,nis10,mor14}. 
\citet{cro10} gave a lower limit on the field strength, $>50$ $\mu$G, over the 
central 400 pc region from a non-thermal radio spectrum. 
By other considerations, it is argued that inferred magnetic field strength 
there is as strong as $\sim$ mG \citep{mor92,fer09}.

If such strong magnetic field is distributed in the bulge, it affects the 
global gas dynamics there \citep[e.g.,][]{sof07}. In fact,
\citet{fuk06} discovered two large molecular loops, loop 1 and loop 2, at a radius of 700-pc from the center, which are triggered by magnetic buoyancy \citep[Parker 
instability; ][]{pak66};  while the main focus in \citet{pak66} 
was aimed at the Galactic disk, he already suggested that the magnetic 
activity might become even more important in the central few 100 pc.
Subsequently, \citet{fuj09,tor10a,tor10b,req10,kud11} presented analyses of more detailed observational properties of these molecular loops, and it was shown that the molecular loops are also well reproduced by 
magneto-hydrodynamical (MHD, hereafter) simulations \citep{sm91,mac09,tak09}. 

\if0 %%% removed by Torii
Recent observation further reveals CO emission at high Galactic latitude, 
$b$, including several filamentary features in addition to diffuse extended 
halo-like CO gas up to 2 degrees in $b$ above and below the CMZ \citep{tor14a}. 
A plausible interpretation is that these high-$b$ features 
also are driven by the Parker instability; although they are filamentary 
without a clear loop-like shape having two foot points, the MHD numerical 
simulations by \citet{mac09} show that magnetic-flotation loops are generally 
not symmetric and can be seen as an open single filament depending on the 
ambient density/field distribution and evolutionary effect. 
Further signature is associated with the double helix nebula (DHN, hereafter) 
toward $(l,b)=(0.0\; {\rm deg.}, 0.8\; {\rm deg.})$ \citep{mor06}, where 
a column of molecular gas of 200-pc height is found to be associated with 
the DHN which is nearly vertical to the plane \citep{eno14,tor14b}. 
Such a feature is explained as created by a magnetic tower formed above and 
below the central black hole according to the preceding numerical simulations 
\citep{kat04,mac09}.

In addition, the importance of magnetic activity in the formation of 
the Fermi bubbles, which are recently found two gigantic 
gamma-ray bubbles originating from the Galactic center region \citep{su10}, 
is also pointed out \citep{car13}.
\fi

In spite of the broad recognition of the strong magnetic field 
in the Galactic center, there is little theoretical work that explores 
the role of the magnetic field in the dynamics of the gas component, 
except a limited number of attempts \citep[e.g.,][]{mac09,ks12,mac13}.  
%Following these observational findings that are supposed to arise from the 
%magnetic activity, 
We naturally expect that the magnetic field may play 
an essential role in the noncircular motion of the gas in the Galactic center 
region, which is the main aim of the present paper. 

In this paper, we investigate the evolution and the role of magnetic field 
in the central region of the Milky Way by a three-dimensional (3D) MHD 
simulation. % that includes the bulge and the disk. 
We do not take into account a bar potential but 
an axisymmetric potential to focus on the role of the magnetic field in 
exciting the noncircular motion of gas.  

\section{Simulation Setup}
\label{sec:setup}
%\if0
\begin{figure}%[h] 
  \begin{center}
    \includegraphics[height=0.36\textheight]{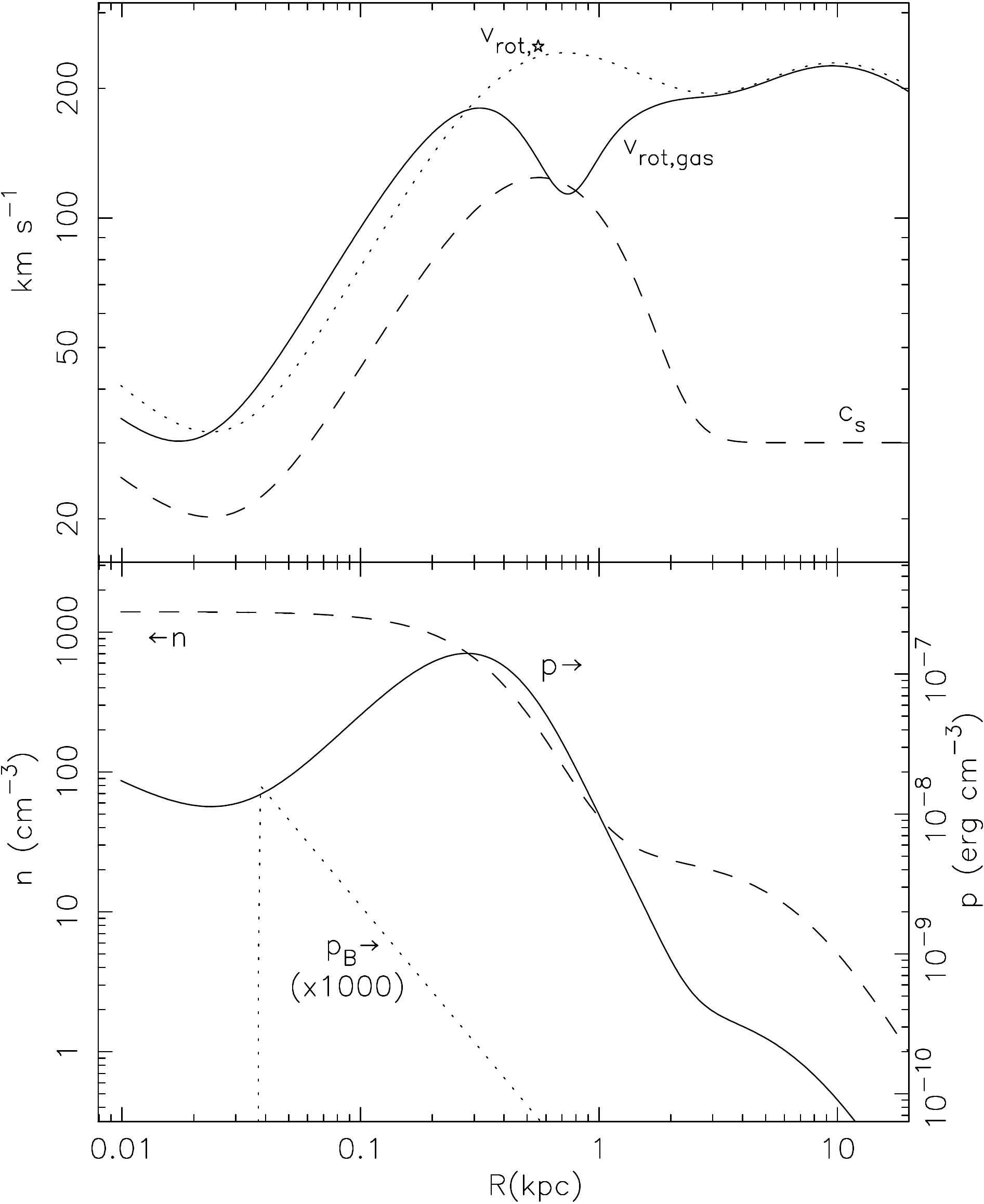}
  \end{center}
\caption{Radial profile of the equilibrium initial condition at the Galactic plane. 
{\it top}: Sound speed (dashed) and initial rotation speed (solid) at the 
Galactic plane. The rotation speed of the stellar component, $v_{\rm rot,\star} = \sqrt{R
\frac{\partial \Phi}{\partial R}}$ (dotted) is also shown for comparison. 
{\it Bottom}: 
Initial density (dashed; left axis), gas pressure (solid; right axis), 
and magnetic pressure, $p_{\rm B}=B^2/8\pi$ (dotted; right axis). 
$p_{\rm B}$ is multiplied by 1000 times to fit in the panel. 
%Subscript ``mid'' stands for
%``midplane'' at $\theta=\pi/2$ and subscript ``suf'' stands for ``surface'' 
%at $\theta = \pi/2\pm\pi/6$
}
\label{fig:init}
\end{figure}
%\fi

We treat the evolution of gas by solving MHD equations under an external 
axisymmetric Galactic gravitational potential. % by stars and dark matter.  
%consisting of three components, the supermassive blackhole (SMBH) at the Galactic center (component $i=1$), the bulge ($i=2$) and the disk ($i=3$).
We consider three components of gravitational sources: The supermassive 
blackhole (SMBH) at the Galactic center (component $i=1$), a stellar 
bulge ($i=2$), and a stellar disk ($i=3$). 
For the SMBH we assume a point mass, $M_1=4.4\times 10^6 M_{\odot}$ 
\citep{gen10}, where $M_{\odot}$ is the solar mass.   
For the $i=2$ and 3 components, we adopt a gravitational potential 
introduced by \citet{mn75} for the bulge and disk components.
%, of which expression can be written in cylindrical coordinates, $(R,\phi,z)$, as 
The gravitation potential that includes these three components is written as 
\begin{equation}
\Phi(R,z)= \sum_{i=1}^{3}\frac{-GM_i}{\sqrt{R^2+(a_i+\sqrt{b_i^2+z^2})^2}},
\label{eq:potential}
\end{equation}
where $R$ and $z$ are cylindrical radius and vertical distance, respectively,  
%usual independent variables of cylindrical coordinates. 
%and  $i=1$, $2$, and $3$ correspond to the supermassive blackhole at the 
%Galactic center, the bulge component and the disk component, respectively.  
%For the bulge and the disk components we adopt the same values for the 
%parameters of the gravitational potential used in \citet{mac09} 
and the adopted values for $M_i$, $a_i$, and $b_i$ are summarized in Table 
\ref{tab:par};
$a_1=b_1=0$ since we assume a point mass for the SMBH,
%with $M_1=4.4\times 10^6 M_{\odot}$, 
and for the bulge and disk components 
we use the same values for $M_i$, $a_i$, and $b_i$ in \citet{mn75} 
(see also \citet{mac09}). 
%: $M_3=2.05\times 10^{10}M_{\odot}$,  
%$M_4=25.47\times 10^{10}M_{\odot}$, $a_3=0$ pc, $a_4=7.258$ pc, 
%$b_3=0.495$ pc, and $b_4=0.52$ pc. 
%The inner bulge ($i=2$) is introduced to account for observed fast 
%rotation near the Galactic center by \citet{sof13}, of which we follow 
%the prescription. 
%The shape of the inner bulge is assumed to be spherical and the mass, 
%$M_{2}(r)$, inside spherical radius, $r (= \sqrt{R^2 + z^2})$, is  
%\begin{equation}
%M_2(r) = M_{2,0} - M_{2,0}\exp(-r/a_2)\left(1+\frac{r}{a_2} + \frac{r^2}{2a_2^2}\right), 
%\end{equation} 
%where $M_{2,0} = 5\times 10^7 M_{\odot}$ and $a_2 = 3.8$ pc. 

%\begin{deluxetable}{cccc}[h]
%\tabletypesize{\scriptsize}
%\tablecaption{Parameters for \citet{mn75} gravitational potential\label{tab:par}
%}
%\tablehead{
%\colhead{Component} 
%& \colhead{$M_i(10^{10}M_{\odot})$} 
%& \colhead{$a_i$(kpc)} 
%& \colhead{$b_i$(kpc)} 
%}
%\startdata
%SMBH 1 & $4.4\times 10^{-4}$ & 0 & 0 \\ 
%Bulge 2 & 2.05 & 0 & 0.495 \\ 
%Disk 3 & 25.47  & 7.258 & 0.52 
%%\hline
%\enddata
%%\tablecomments{}
%\end{deluxetable}

\begin{table}%[h]
\begin{tabular}{cccc}
\hline
 & $M_i(10^{10}M_{\odot})$ & $a_i$(kpc) & $b_i$(kpc)\\
\hline
\hline
SMBH 1 & $4.4\times 10^{-4}$ & 0 & 0 \\ 
\hline
Bulge 2 & 2.05 & 0 & 0.495 \\ 
\hline
Disk 3 & 25.47  & 7.258 & 0.52 \\
\hline
\end{tabular}
\caption{Parameters for the gravitational potential. 
The parameters for the super-massive black hole (SMBH) are from \citet{gen10}, 
and the parameters for the bulge and disk components are from \citet{mn75}.
\label{tab:par}
}
\end{table}

Assuming an equation of state for ideal gas, the gas pressure, 
\begin{equation}
p = \rho c_{\rm s}^2,
\label{eq:eos}
\end{equation}
where $\rho$ is gas density and $c_{\rm s}(\propto \sqrt{T})$ is sound speed. 
We consider the spatial dependence of temperature ($\propto c_{\rm s}^2$) 
but assume that it is kept constant with time in each grid cell; the gas 
is assumed to be locally isothermal.
Since we treat the Galactic bulge and disk 
by global simulation, the ``sound speed'' here more or less reflects 
the velocity dispersion of gas clouds rather than the actual 
temperature of each cloud.  
In the disk region, we assume $c_{\rm s}$ is comparable to the 
observed velocity dispersion \citep[e.g.,][]{bov12},
\begin{equation}
c_{\rm s,disk} = 30 {\rm km\; s^{-1}},
\end{equation}
whereas this value is much smaller than the rotation speed $\sim 200$ km s$^{-1}$.
In the bulge region, however, large velocity dispersion, which is probably 
because of the noncircular motion, is obtained \citep[e.g.,][]{ken92}. 
To mimic this larger velocity dispersion, we adopt 
\begin{equation}
c_{\rm s,bulge} = 0.6 v_{\rm rot,\star} = 0.6 \sqrt{R \frac{\partial \Phi}{\partial R}},
\end{equation}
where $v_{\rm rot,\star}$ is the rotation speed of the stellar component. 

We smoothly connect these two components to fix the radial dependence: 
\begin{equation}
\hspace{-0.5cm}c_{\rm s}^2 = c_{\rm s,bulge}^2 \left[1 - 
\tanh\left(\frac{R-R_{\rm b}}{\Delta R_{\rm b}}\right)\right]  
+ c_{\rm s,disk}^2\left[1 + 
\tanh\left(\frac{R-R_{\rm b}}{\Delta R_{\rm b}}\right)\right], 
\end{equation}
where we assume the boundary between the bulge and the disk, $R_{\rm b} = 1$ kpc, 
with a smoothing width, $\Delta R_{\rm b} = 0.8$ kpc. 

%We set up the initial density profile derived from the Poisson equation:
%\begin{eqnarray}
%& &\rho(R,z) = \frac{\Delta \Phi}{4\pi G}\\
%&=& \frac{1}{4\pi}\sum_{i=2,3}\frac{M_i b_i^2 
%\left[a_i R^2 + \left(a_i + 3\sqrt{b_i^2 + z^2}\right)
%\left(a_i + \sqrt{b_i^2 + z^2}\right)^2\right]}{\left[R^2 + 
%\left(a_i + \sqrt{b_i^2 + z^2}\right)^2\right]^{5/2}
%  \left(b_i^2 + z^2\right)^{3/2}}, \nonumber
%\end{eqnarray}
%where we sum up the only $i=2,3$ components since the contribution from the 
%point-mass SMBH ($i=1$) is zero. 

We start our simulation from an equilibrium configuration, in which the 
gravity is balanced with the pressure-gradient and centrifugal forces: 
%(see Appendix for detail) :
\begin{equation}
-\frac{1}{\rho}\frac{\partial p}{\partial R} + R \Omega^2 -\frac{\partial \Phi} 
{\partial R} = 0,
\label{eq:hydstr}
\end{equation}
and
\begin{equation}
-\frac{1}{\rho}\frac{\partial p}{\partial z} -\frac{\partial \Phi} 
{\partial z} = 0. 
\label{eq:hydstz}
\end{equation}
%where $\Omega$ changes with time after the simulation starts.
We assume the initial gas density at the Galactic plane is proportional to the 
stellar density that fixes the gravitational potential, namely 
$\rho \propto \frac{\Delta \Phi}{4\pi G}$.
Then, we determine the distribution of the initial density and  
rotation frequency, $\Omega$,  to satisfy Equations (\ref{eq:hydstr}) and 
(\ref{eq:hydstz}).

Figure \ref{fig:init} presents the radial distribution of the initial 
equilibrium profile at the Galactic plane. The rotation speed of the gas 
component shows a bump near $R=0.5$ kpc, associated with a peak 
of the sound speed there (top panel). 
%In order to fulfill the force balance in the vertical direction, the large mass 
%contained in the small bulge region is necessarily supported by the gas 
%pressure. This requires high temperature (large $c_{\rm S}$) 
%in the central region, $R<1$ kpc, which also affects the radial force balance. 
In this region, because of the high temperature, the outward pressure-gradient 
force is not negligible, compared to the centrifugal force; the inward gravity 
is balanced by both the pressure-gradient force and the centrifugal force. 
To satisfy this radial force balance, the rotation speed should be kept 
considerably smaller than the rotation speed of the stellar component there 
to suppress the centrifugal force. 
This equilibrium configuration plays an important role in the 
amplification of the magnetic field as will be discussed later. 
It should also be noted that the mass distribution of stars 
in the inner $\sim$ 0.1 kpc has recently been obtained \citep{sof13,fel14}, 
which shows the existence of a certain amount of mass ($\sim 10^8M_{\odot}$) 
there.
We need to incorporate this contribution in future studies, which would give 
moderately faster rotation speed in the inner bulge.

The bottom panel of Figure \ref{fig:init} presents the initial density, 
the gas pressure, and the magnetic pressure. The density is expressed in units 
of number density, $n$ cm$^{-3}=\rho /\mu m_{\rm H}$, with mean molecular 
weight, $\mu=1.2$, where $m_{\rm H}$ is the mass of a hydrogen atom. 
In this work we adopt the one-fluid approximation and do not distinguish ions, 
neutral atoms, and molecules. $n$ with $\mu=1.2$ approximately corresponds to
number density in units of hydrogen atoms. For rough estimates of molecular 
number density, which is dominated by H$_2$, in dense clouds, we can use 
$n_{\rm H_2}\approx 0.5 n$.
Initially we set up weak vertical magnetic field with,
\begin{equation}
B_z = 0.71\mu{\rm G}\left(\frac{R}{1\;{\rm kpc}}\right)^{-1}, 
\end{equation}
in the region of $R>0.035$ kpc.
A reason why we start from the weak magnetic field is that we would 
like to avoid the direct effect of the initial magnetic field on 
the dynamics of the gas. 
The initial magnetic pressure, $p_{\rm B}=B^2/8\pi$, is 
%$10^{-5}-10^{-4}$ times 
less than $<10^{-3}$ times the gas pressure 
%except in $0.035<R<0.1$ kpc 
as shown in the bottom panel of Figure \ref{fig:init}, namely the plasma 
$\beta\equiv p/p_{\rm B}>10^{3}$. 

%\begin{deluxetable}{cccccc}[h]
%\tabletypesize{\scriptsize}
%\tablecaption{Simulation Box \& Numerical Resolution\label{tab:box}
%}
%\tablehead{
%\colhead{$r$(kpc)} 
%& \colhead{$\theta$} 
%& \colhead{$\phi$} 
%& \colhead{$N_r$} 
%& \colhead{$N_{\theta}$} 
%& \colhead{$N_{\phi}$}
%}
%\startdata
%(0.01,60) & ($\pi/6,5\pi/6$) & ($-\pi,\pi$) & 
%384  & 128 & 256 
%%\hline
%\enddata
%%\tablecomments{}
%\end{deluxetable}

\begin{table}%[h]
\begin{tabular}{cccccc}
\hline
$r$(kpc) & $\theta$ & $\phi$ & $N_r$ & $N_{\theta}$ & $N_{\phi}$ \\
\hline
(0.01,60) & ($\pi/6,5\pi/6$) & ($-\pi,\pi$) & 
384  & 128 & 256 \\
\hline
\end{tabular}
\caption{Simulation Box \& Numerical Resolution\label{tab:box}}
\end{table}

%\if0
\begin{figure}
  \begin{center}
    \includegraphics[width=0.4\textwidth]{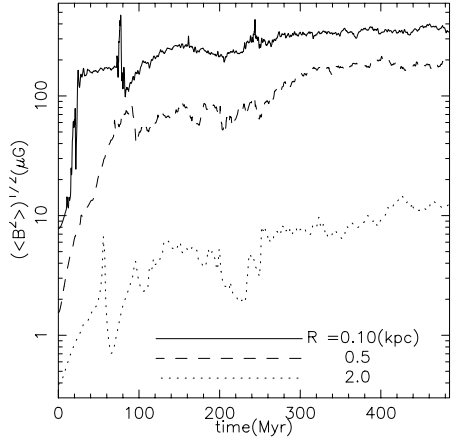}
  \end{center}
  \caption{
    Time evolution of the magnetic field strength averaged over 
    the azimuthal and vertical directions (see {\it text}) at different radial 
    locations, $R=0.1$ kpc (solid), 0.5 kpc (dashed), and 2.0 kpc (dotted). 
    \label{fig:tevol}
  }
\end{figure}

\begin{figure}%[h]
  \begin{center}
    \includegraphics[height=0.3\textheight]{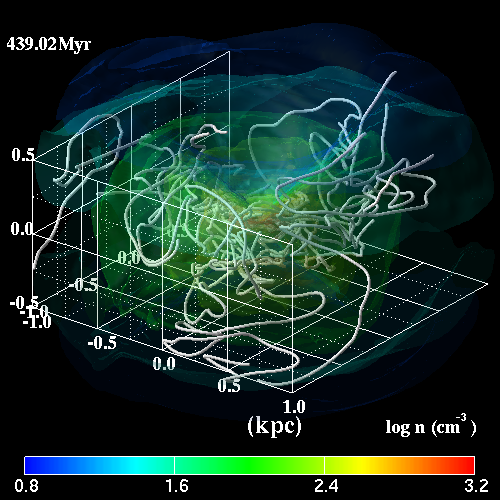}\\
    \vspace{0.1cm}
    \includegraphics[height=0.2\textheight]{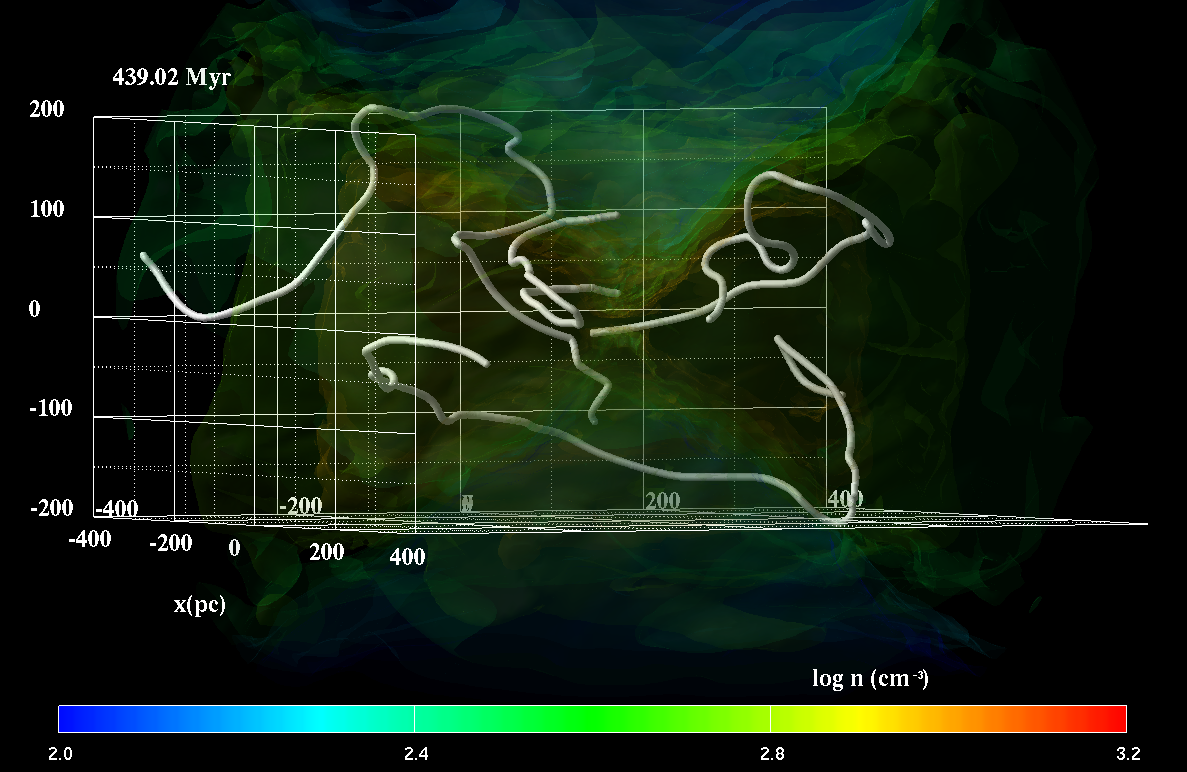}
%    \includegraphics[height=0.3\textheight]{../worksphmpi/snp28_t27-10pi_2.png}\\
%    \vspace{0.1cm}
%    \includegraphics[height=0.2\textheight]{../worksphmpi/snp28_t27-10pi_1.png}
  \end{center}
  \caption{Snapshot views of the bulge region ($R<1$ kpc) of the simulation 
    at $t=439.02$ Myr (upper panel). 
    Colours indicate density in units of number 
    density, $n$ cm$^{-3}$, in logarithmic scale, white lines denote 
    magnetic field lines. %, and arrows indicate velocity. 
    In the lower panel the central region is zoomed in and seen from a nearly 
    edge-on angle to show $\frown$-shaped field lines which are typical for 
    Parker instability.
    The movies are available (see text and footnote). 
    \label{fig:snp1}
  }
\end{figure}
%\fi

%We perform numerical simulations 
We update $\rho$, $\mbf{v}$, and $\mbf{B}$ with time by solving ideal 
MHD equations, 
\begin{equation}
\frac{\partial \rho}{\partial t} + \mbf{\nabla}\cdot(\mbf{\rho v}) = 0,
\end{equation}
\begin{equation}
\rho\frac{\partial \mbf{v}}{\partial t} = -\rho(\mbf{v\cdot\nabla}){\mbf v}
-\mbf{\nabla}\left(p + \frac{B^2}{8\pi}\right) 
+ \left(\frac{\mbf{B}}{4\pi}\cdot \mbf{\nabla}\right)\mbf{B}
%- \rho \frac{GM}{r^2},
-\rho \mbf{\nabla}\Phi,
\label{eq:mom}
\end{equation}
and
\begin{equation}
\frac{\partial \mbf{B}}{\partial t} = \mbf{\nabla \times (v\times B)}, 
\end{equation}
under the fixed gravitational potential (Equation \ref{eq:potential}). 
The numerical scheme we adopt is the second-order Godunov-CMoCCT 
method \citep{san99}, in which we solve nonlinear Riemann problems with 
magnetic pressure at cell boundaries for compressive waves and adopt 
the consistent method of characteristics (CMoC) for the evolution of 
magnetic fields \citep{cl96,sn92} under the constrained
transport (CT) scheme \citep{eh88}.
Since we assume the locally isothermal equation of state (Equation 
\ref{eq:eos}), we do not solve an energy equation.  
We perform our simulation in {\it spherical coordinates}, 
$(r,\theta,\phi)$, instead of cylindrical 
coordinate, $(R,\phi,z)$, to resolve the central region by fine-scale grids. 
The size of the simulation box and the number of the grid points are 
summarized in Table \ref{tab:box}. 
In the $\theta$ direction, the simulation domain covers $\pm \pi/3$ 
from the Galactic plane; it does not cover the regions near the poles, and 
we need to set up an appropriate condition for the $\theta$ boundaries, 
which is described below.
The size of the radial grid, $\Delta r$, is enlarged in proportion to $r$.  
We analyze the numerical data mainly in the cylindrical coordinates 
by converting the primitive data in the spherical coordinates.

We adopt the same boundary condition as in \citet{si14}. We prescribe the 
outgoing condition for both mass and waves at the $\theta$ boundaries, which 
was originally developed for simulations for the solar wind \citep{si05,si06}, 
and the accretion condition at both the outer and inner $r$ boundaries. 
Because of the outgoing boundaries at the $\theta$ surfaces, the mass 
rapidly streams out by vertical outflows driven by effective magneto-turbulent 
pressure \citep{si09,si14}. Therefore, the total mass in the 
simulation box does not conserve but decreases with time.
This process probably overestimates 
the mass loss especially in the bulge region because our simulation box 
does not cover up to the sufficiently higher altitude \citep{suz10,fro13}. 
In order to compensate the rapid mass decrease in the bulge, we use very 
large ``density floor'', 
$\rho_{\rm fl}$ of 0.8 times the local initial value, $\rho _{\rm init}$; 
if density becomes less than the floor value, $\rho_{\rm fl} 
= 0.8\rho_{\rm init}$, at each location, we recover it to $\rho_{\rm fl}$.   
The density floor is utilized in the only regions near the $\theta$ surfaces 
where the mass is lost by vertical outflows.
The density there is lower than that near the Galactic plane, and the effect of 
mass gain by this treatment is small and it only partially compensates 
the mass lost by vertical outflows. 

When we carry out the simulation, all the physical variables are 
non-dimensionalized by unit mass, $m_{\rm unit}=10^{10}M_{\odot}$, unit length, 
$l_{\rm unit} = 1$ kpc, and unit time, $t_{\rm unit} = \left(\frac{Gm_{\rm unit}}
{l_{\rm unit}^3}\right)^{-1/2}=5.15$ Myr. We perform the simulation up to $t
=30\pi t_{\rm unit} = 485.21$ Myr.

\section{Results}
%\if0
\begin{figure*}%[t]
  \begin{center}
    \includegraphics[height=0.3\textheight]{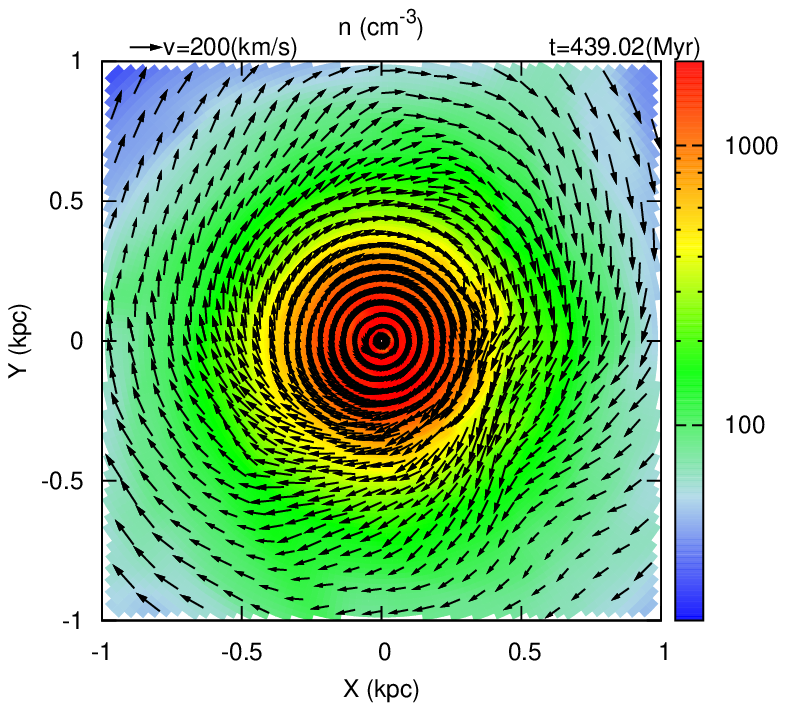}\hspace{-3.6cm}
    \includegraphics[height=0.3\textheight]{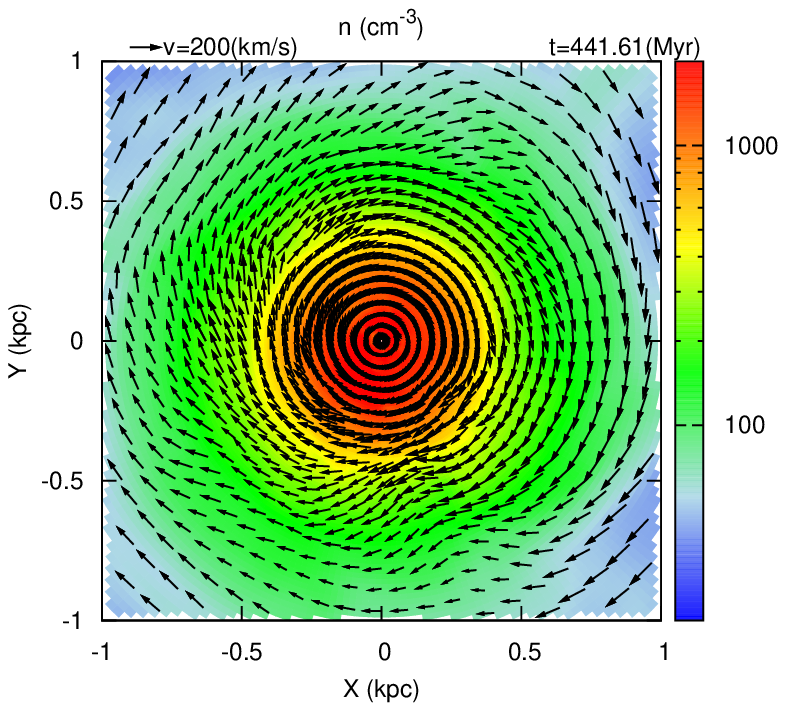}
  \end{center}
  \caption{Face-on views of density in units of $n$ cm$^{-3}$ ({\it colour}) and velocity field 
    ({\it arrows}) at the Galactic plane at different times, $t=439.02$ Myr 
    ({\it left}) and 441.61 Myr ({\it right}).
    \label{fig:faceon1}
  }
\end{figure*}

\begin{figure*}%[t]
  \begin{center}
    \includegraphics[height=0.24\textheight]{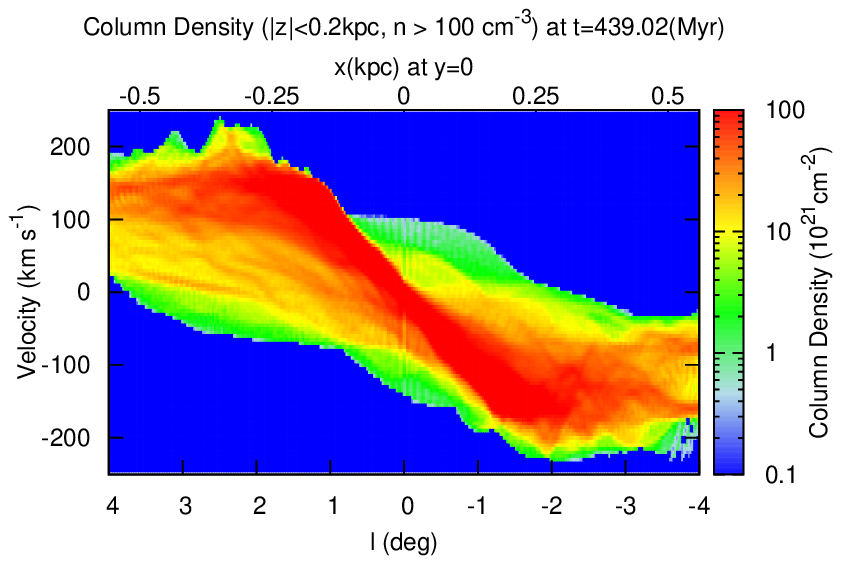}
    \includegraphics[height=0.24\textheight]{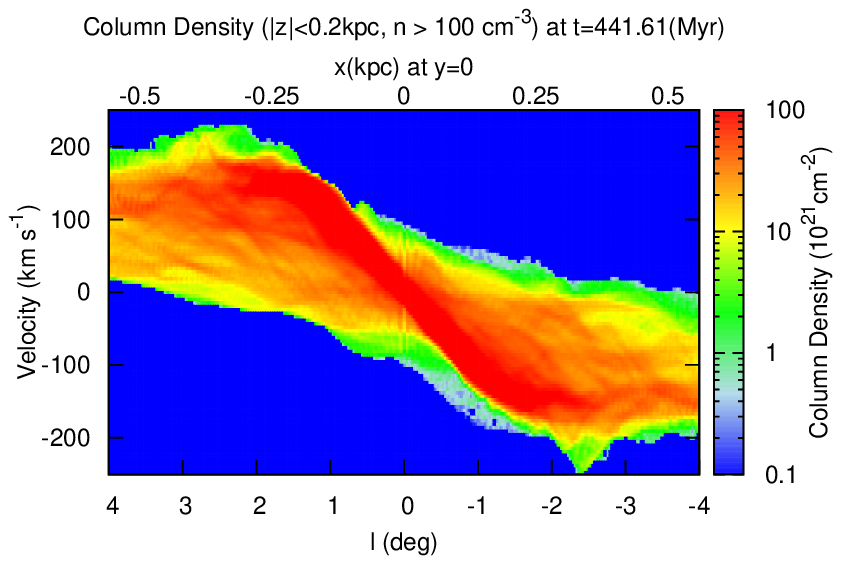}
  \end{center}
  \vspace{-0.6cm}
  \caption{$l$--$v$ diagrams  at different times, $t=439.02$ Myr 
({\it left}) and 441.61 Myr ({\it right}). The grid points with high-density 
regions, $n>100$ cm$^{-3}$, in $|z|<0.2$ kpc 
%and $R<5$ kpc 
are used to derive the column density in units of 
$10^{21}$cm$^{-2}$ ({\it colour}) by 
integrating $n$ along the direction of line of sight.  
%For the bottom panels, the only high-density regions with 
%$N_{\rm H}>500$ cm$^{-3}$ are used for the integration to focus on the central 
%region. 
In the horizontal axis, both Galactic longitude, $l$ degree, ({\it bottom 
axis}) and the corresponding transverse distance, $x$ kpc, 
at $y=0$ kpc ({\it top axis}) are shown, where the solar system is located 
at $(x,y)=(0,-8)$ kpc.  
\label{fig:p-v1}
}
\end{figure*}

\begin{figure*}%[t]
  \begin{center}
    \includegraphics[height=0.24\textheight]{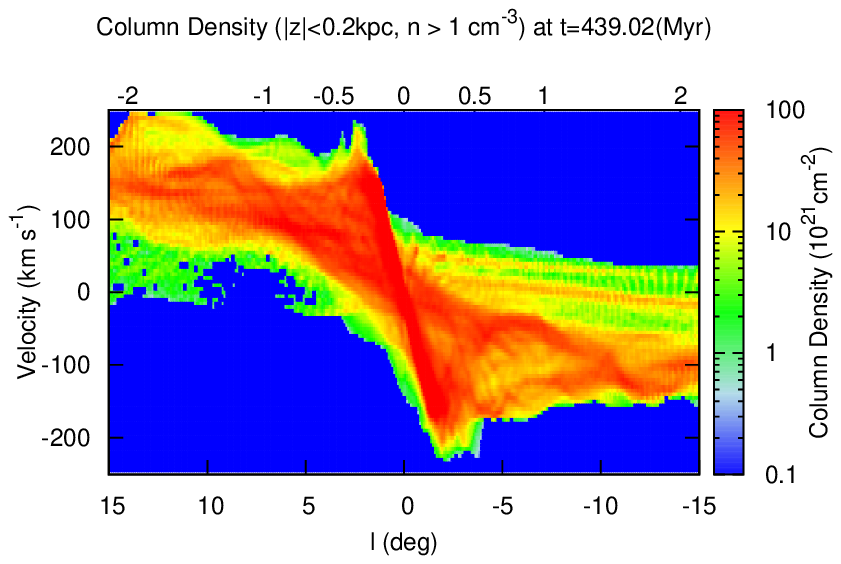}
    \includegraphics[height=0.24\textheight]{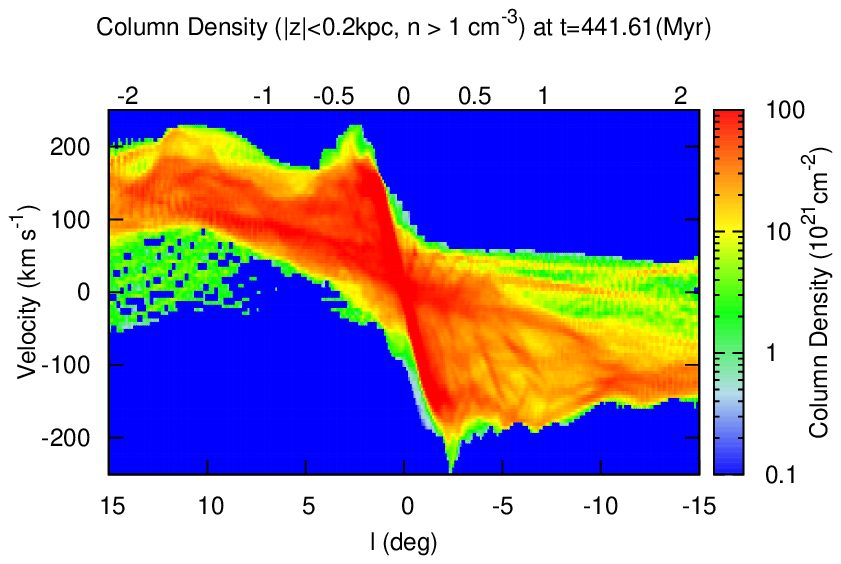}
  \end{center}
  \vspace{-0.6cm}
  \caption{Same as Figure \ref{fig:p-v1} but for a large $l$ range. 
    Regions with smaller density, $n>1$ cm$^{-3}$, are considered to derive 
    the column density than for Figure \ref{fig:p-v1}. 
\label{fig:p-v2}
}
\end{figure*}
%\fi

\subsection{Overview}

After the simulation starts, the amplification of the weak vertical seed
magnetic field is initially triggered by magneto-rotational instability 
\citep[MRI, hereafter; ][]{vel59,cha61,bh91} and the vertical shear  
of the initial rotation profile \citep{si14,mp14} to generate the radial 
and toroidal components. 
%{\color{red}
The wavelength of the most unstable mode of the MRI, which is  
captured by $\sim$ one grid scale, is slightly underresolved at first 
because the initial field strength is weak. 
However, the wavelength of the most unstable mode, which is proportional 
to the field strength, becomes longer with the amplification of magnetic 
field, and it can be well resolved in the saturated state.%}
The toroidal component is further amplified by the winding due to 
the radial differential rotation. Furthermore, magnetic buoyancy \citep{pak66} 
also plays a role in amplifying the vertical component \citep{nis06,mac13}.

Figure \ref{fig:tevol} presents the time evolution of the magnetic 
field strength averaged over azimuthal ($\phi$) and vertical ($z$) directions, 
\begin{equation}
\sqrt{\langle B^2\rangle_{\phi,z}}= \frac{\sqrt{\int_{-\pi}^{\pi}d\phi
\int_{z_1}^{z_2}dz (B_R^2+B_{\phi}^2+B_z^2)}}{{2\pi (z_2-z_1)}}, 
\end{equation}
at different radial locations, $R=0.1$ kpc (solid), 0.5 kpc (dashed), and 2.0 
kpc (dotted), where the vertical integration is taken from $z_1=-1$ to 
$z_2=1$ kpc. The growth of magnetic field is faster at smaller $R$ because 
the growth rate of the MRI and the winding is scaled by rotation frequency, 
$\Omega$ (see \S \ref{sec:oncm}), which decreases with $R$. Figure 
\ref{fig:tevol} shows that the field strength saturates after $t\gtrsim 400$ 
Myr at $R=0.1$ and 0.5 kpc in the bulge region, while it is still gradually 
growing at $R=2$ kpc until the end of the simulation, $t=30 \pi t_{\rm unit} 
=485.21$ Myr.  

Figure \ref{fig:snp1} presents 3D snapshots
\begin{lrbox}\myVerb
  \scriptsize{\verb|www.ta.phys.nagoya-u.ac.jp/stakeru/research/galaxypot/glbdsk28inv_lagp2.mp4| }
\end{lrbox}
\footnote{Movie of the simulation is available at \\ \usebox\myVerb}
%\footnote{
%\\A movie showing a larger region up to $R=2$ kpc is also at \\
%    www.ta.phys.nagoya-u.ac.jp/stakeru/research/galaxypot\\/glbdsk26inv\_lagp1.mp4
%}
of the bulge region, $R<1$ kpc, after the MHD turbulence is well 
developed at $t = 439.02$ Myr. 
%Here, the density is shown in units of number density of hydrogen atoms, 
The figure exhibits 
%turbulent and highly structured nature of the gas and the magnetic field 
turbulent gas and complicatedly tangled magnetic field lines
in the bulge region.  
%The magnetic field seems to be mostly dominated by the toroidal component, 
One can see turbulent poloidal field lines excited by MRI and Parker 
instability, whereas the toroidal component slightly dominates the poloidal 
(radial and vertical) component as will be discussed in \S \ref{sec:oncm}. 
In the lower panel that zooms in the central region, we pick up 
$\frown$-shaped field lines excited by Parker instability, 
which are actually observed in the Galactic center region \citep{fuk06}. 
%Rotating motion is seen in the velocity field,  
%which is also disturbed by radial and vertical fluctuations. 
%{\color{red} 
This turbulent state is sustained in a quasi-steady manner after 
the field strength is saturated. In Figure \ref{fig:tevol} we only show the 
evolution of the average field strength, and then one cannot see  
evolutionary properties of the magnetic configuration and turbulence. 
The movie of Figure \ref{fig:snp1} (see footnote for the link) demonstrates 
that the MRI and Parker instability continuously drive non-axysmmetric complex 
magnetic configuration rather than developing coherent axisymmetric structure. 
Therefore, the system is in a quasi-steady saturated state 
in terms of magnetic turbulence, after the saturation of the field strength.   
%}

Figure \ref{fig:faceon1} shows the velocity field with the density, $n$, at 
the Galactic plane ($\theta=\pi/2$) at different times 
\begin{lrbox}\myVerb
  \scriptsize{\verb|www.ta.phys.nagoya-u.ac.jp/stakeru/research/galaxypot/faceon28inv_4.gif| }
\end{lrbox}
\footnote{Movie is available at \\ \usebox\myVerb};
the left panel presents the result at the same time as 
in Figure \ref{fig:snp1}, and the right panel shows the result at 2.6 Myr later 
than the left panel, where this time difference ($=2.6$ Myr) roughly 
corresponds to $\approx 1/3$ of the rotation time at $R=0.1$ kpc. 
These two panels show that radial motion is stochastically excited 
particularly around $R\approx 0.5$ kpc. Outward motion ($v_{R}>0$) appears 
to dominate, whereas one can also recognize inflows ($v_{R}<0$) in some 
regions. 
These behaviors are a result of the magnetic field as will be discussed 
in more detail later (\S \ref{sec:oncm}).

\subsection{$l$--$v$ diagrams}
Figure \ref{fig:p-v1} shows Galactic longitude--velocity ($l-v$) diagrams 
observed from the LSR at different times that correspond to 
Figure \ref{fig:faceon1}. 
The solar system is assumed to rotate with 240 km s$^{-1}$ in the clockwise 
direction at $R=8$ kpc \citep{hon12} and $\phi=-\pi/2$, namely 
$(v_{x},v_y,v_z)=(-240,0,0)$ km s$^{-1}$ at $(x,y,z) = (0,-8,0)$ kpc 
in the Cartesian coordinates. The contours indicate the column density 
in units of $10^{21}$cm$^{-2}$, integrated along the 
direction of line of sight.
%To construct the $l-v$ diagrams in the bottom panels, 
We use the grid points with high-density regions where $n>100$ cm$^{-3}$, 
because the high-density regions more or less trace the cool molecular gas, 
although our simulation treats one-fluid gas and does not distinguish 
molecules, neutral atoms, or ions.

Parallelogram-like shapes are observed in the central region as a consequence 
of the excited radial motion shown in Figure \ref{fig:faceon1}.  
%The bottom panels further exhibit that 
Comparing the two panels, the shape changes with time 
\begin{lrbox}\myVerb
  \scriptsize{\verb|www.ta.phys.nagoya-u.ac.jp/stakeru/research/galaxypot/pv-anim28inv_hr_2.gif| }
\end{lrbox}
\footnote{Movie is available at \\ \usebox\myVerb}
on account of the time-dependent nature of the radial flows. 
Moreover, one can see that the shape is not symmetric because the distribution 
of the excited radial flows is not axisymmetric or bisymmetric. 
Interestingly, the parallelogram shape obtained in the Milky way is not 
symmetric \citep[][then, the shape is not a ``parallelogram'' in a strict 
sense]{dam01,tor10a}, which cannot be explained by a simple bar potential.

Figure \ref{fig:p-v2} presents $l$--$v$ diagrams for a larger 
$|l| < 15$ degree. We use regions with smaller density, $n>1$ cm$^{-3}$, 
to calculate surface density than for Figure \ref{fig:p-v1} 
in order to take into account the outer region where the density is smaller. 
These diagrams also show non-symmetric structure in the $l$--$v$ plane, which 
indicates that non-axisymmetric gas distributions are extended to larger $R$.

%\if0
\begin{figure}%[h]
  \begin{center}
    \includegraphics[height=0.3\textheight]{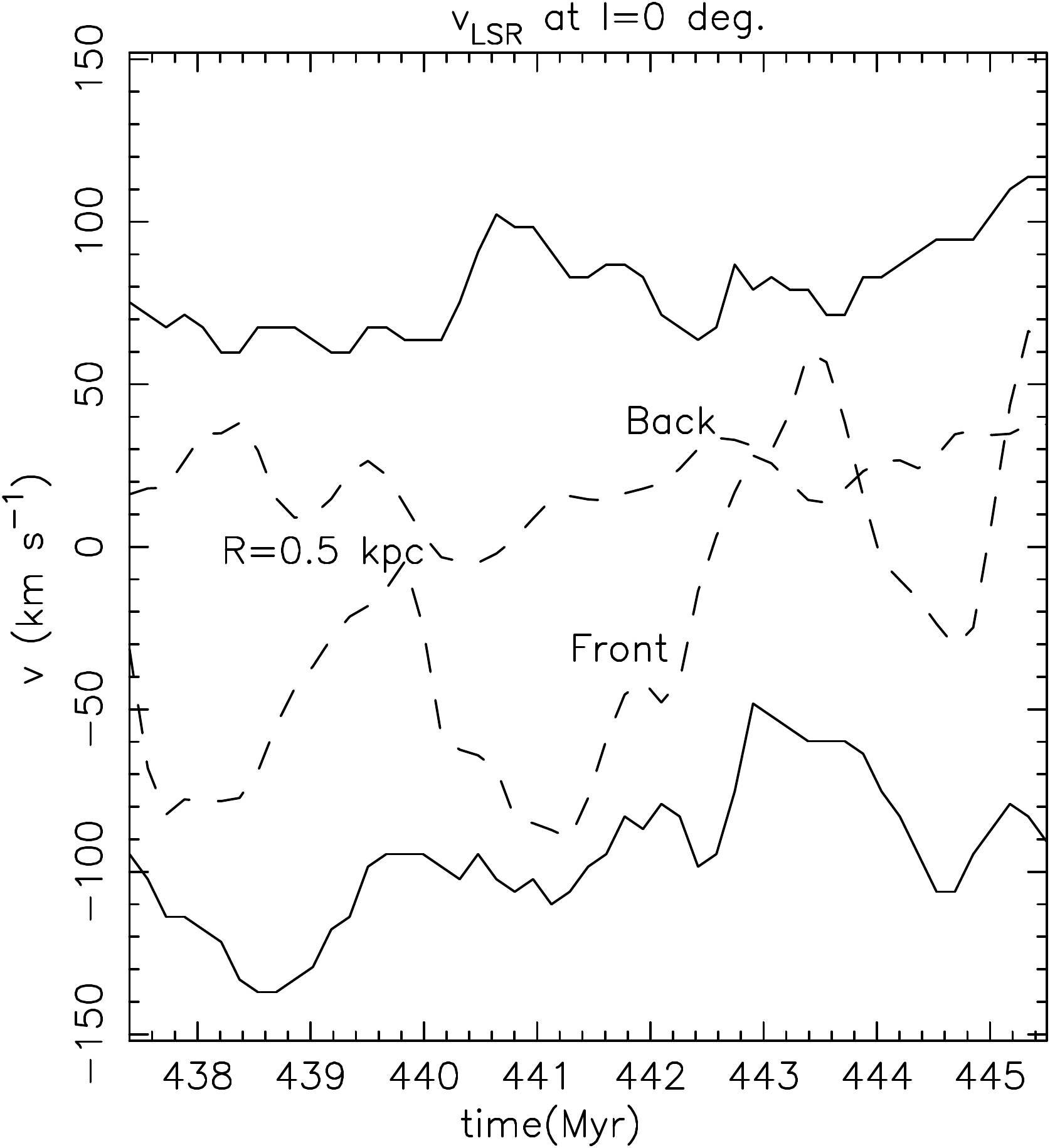}\\    
    \includegraphics[height=0.3\textheight]{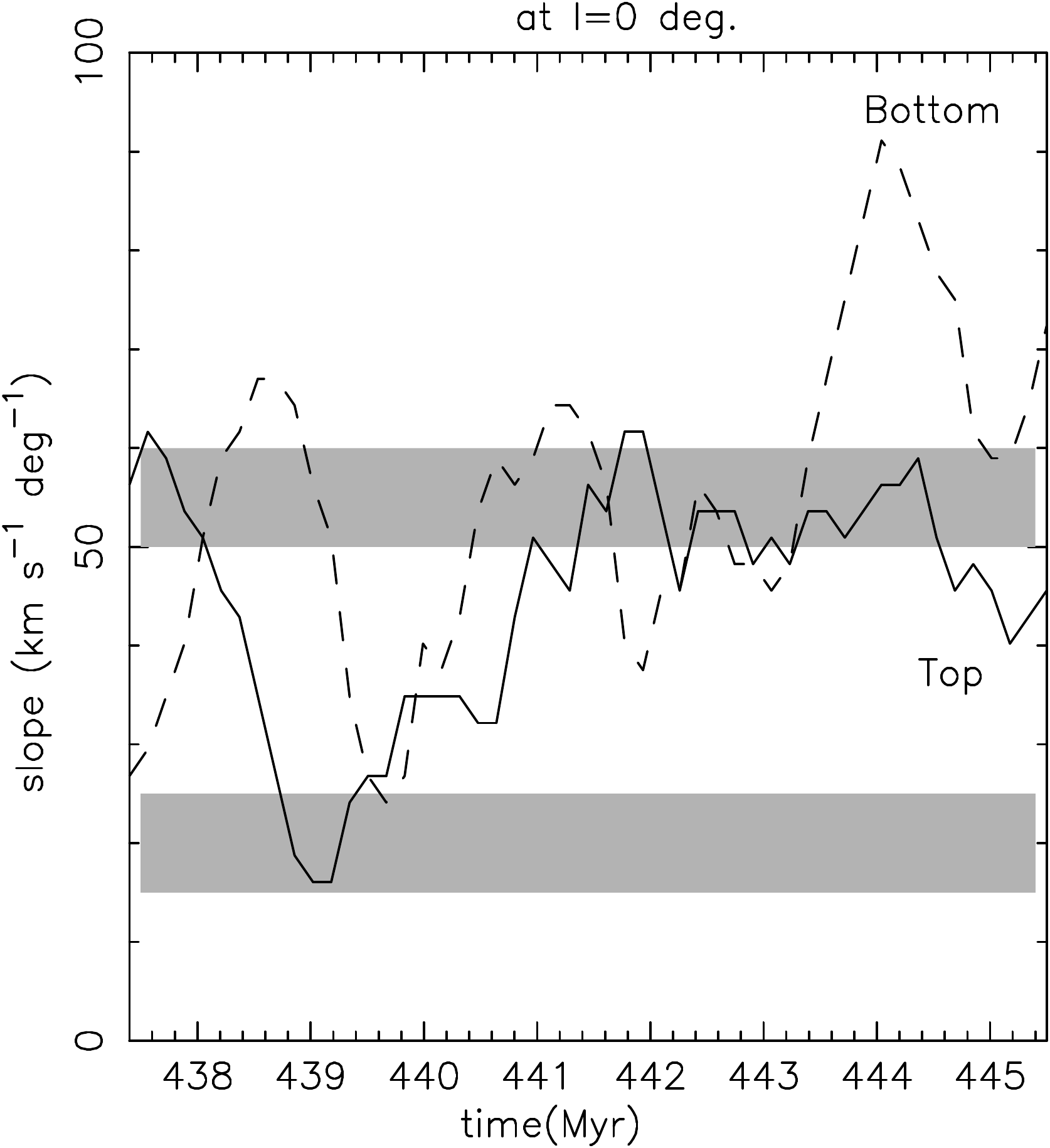}
  \end{center}
  \caption{{\it top:} Time evolution of heliocentric velocities at Galactic 
    longitude, $l=0$ degree The solid lines indicate the maximum and minimum 
    velocities that give column density, $\Sigma > 0.5\times 10^{21}$cm$^{-2}$. 
    The dashed lines trace 
    the radial motion of the front-side ($\phi = -\pi/2$) and back-side 
    ($\phi = \pi/2$) at $R=0.5$ kpc ($(x,y)=(0,\mp 0.5)$ kpc in the Cartesian 
    coordinates).
    {\it bottom:} Time evolution of the slope of the 
      ``parallelogram" near $l=0$ degree. The solid and dashed lines indicate 
      the slope of the top and bottom sides, respectively. The slope is 
      derived from the average between $l = \pm 0.72$ degrees, which 
      corresponds to $\approx \pm 100$ pc around the Galactic center. 
      The shaded regions indicate the observational values, the region 
      centered at 20 km s$^{-1}$deg$^{-1}$ for the top side of the 
      parallelogram and the region centered at 55 km s$^{-1}$deg$^{-1}$ for 
      the bottom side. 
    \label{fig:p-v_tdp}
  }
\end{figure}

\begin{figure}%[h]
  \begin{center}
    \includegraphics[height=0.4\textheight]{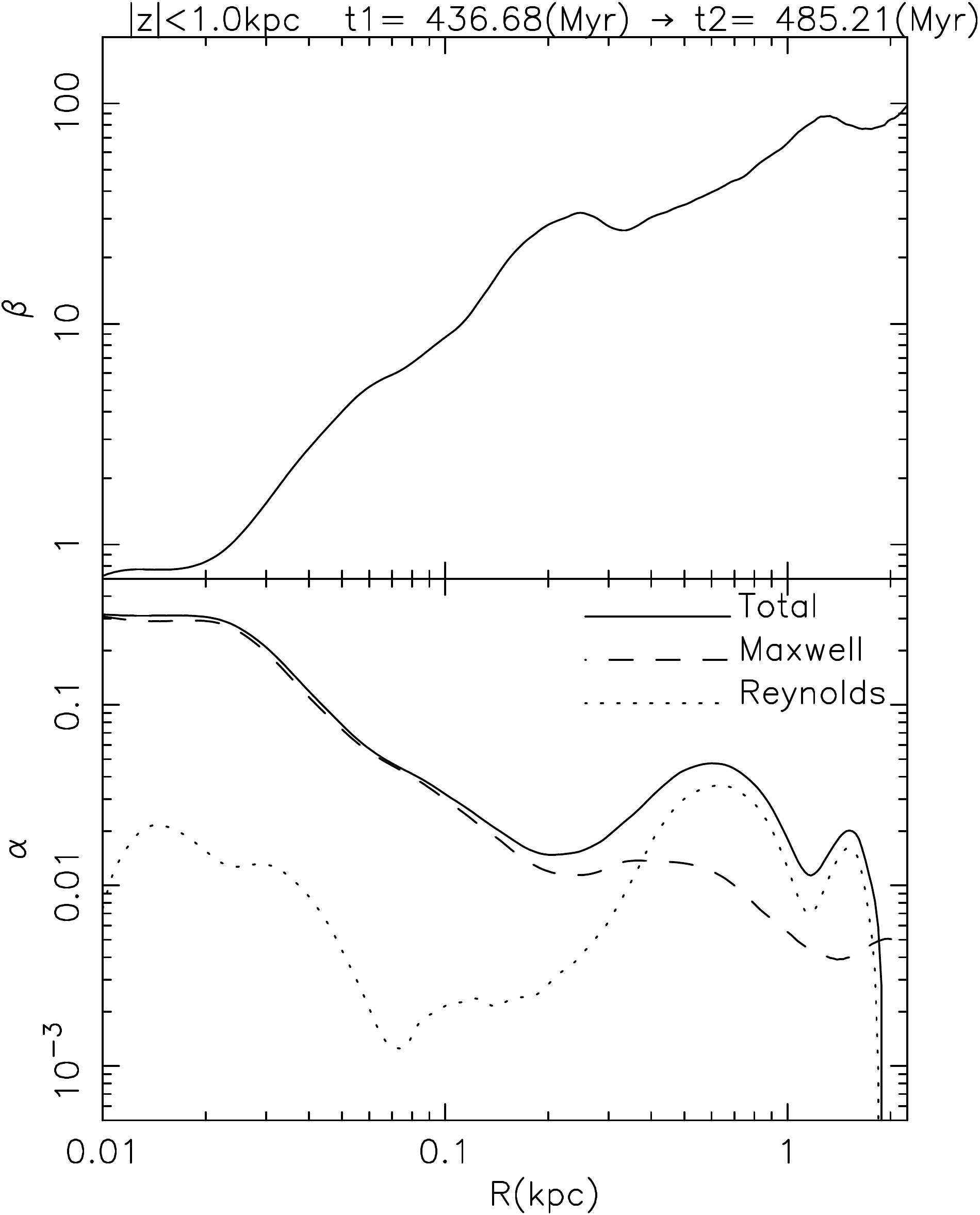}
  \end{center}
  \caption{Radial distribution of the plasma $\beta$ ({\it top}) and the 
    $\alpha$ value ({\it bottom}; solid line) which is the sum of the Maxwell 
    (dashed line) and Reynolds (dotted line) stresses. The data are averaged 
    over the azimuthal direction in full rotation ($2\pi$), over the vertical 
    direction within $|z|<1$ kpc, and the time $t=436.68$--$485.21$ Myr.
    \label{fig:tave1}
}
\end{figure}

\begin{figure}%[h]
  \begin{center}
    \includegraphics[height=0.5\textheight]{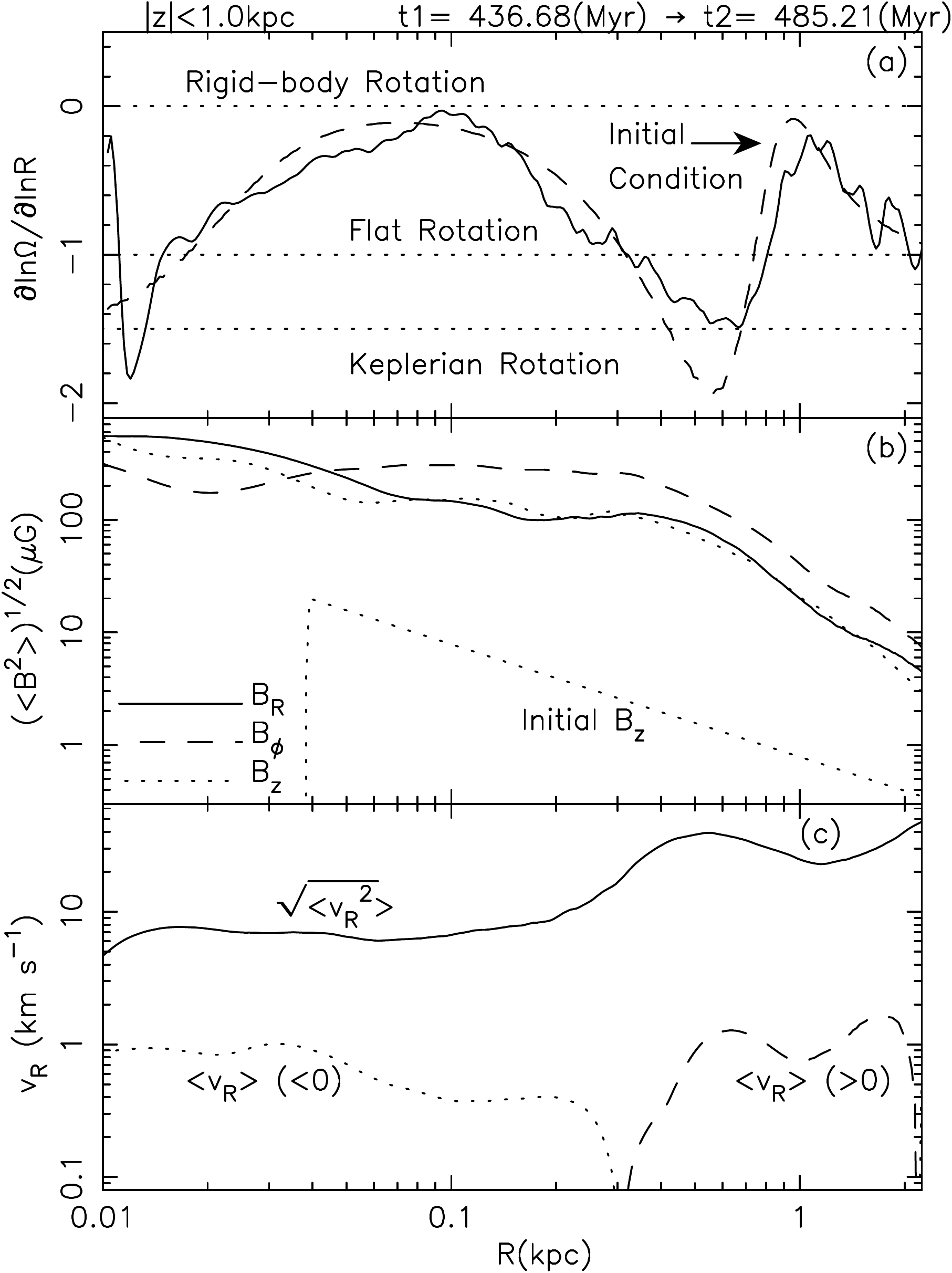}
  \end{center}
  \caption{Radial distribution of various quantities.  
    The data are averaged over the azimuthal direction in full 
    rotation ($2\pi$), over the vertical direction within $|z|<1$ kpc (except 
    for panel {\bf (a)}), 
    and the time $t=436.68$--485.21 Myr.
%    except $\partial\ln\Omega/\partial\ln R$ in the bottom-right panel 
    (see text). 
    {\bf (a)}: The strength of differential rotation, 
    $\partial\ln\Omega/\partial\ln R$ at the Galactic plane (solid line); 
    The average is not taken over the $z$ component but over the $\phi$ 
    component and time. The dashed line represents the initial condition. 
    Rigid-body rotation  
    ($\partial\ln\Omega/\partial\ln R = 0$), flat rotation ($=-1$), 
    and Keplerian rotation ($=-3/2$) are shown by dotted lines for 
    reference. 
    {\bf (b)}: $R$ (solid), $\phi$(dashed), and 
    $z$ (dotted) components of root-mean squared (rms) magnetic field 
    in units of $\mu$G. The dotted line with a constant slope denotes 
    the initial vertical magnetic field. 
%    {\bf (c)}:  Each component of rms velocity.
%    For the $\phi$ component (dashed) the mean rotation is subtracted to 
%    extract the fluctuation component (Equation \ref{eq:dvphi}). 
    {\bf (c)}: Root mean squared radial velocity, $\sqrt{\langle v_{R}^2 
      \rangle}$ (solid), and mean radial velocity, $\langle v_R\rangle$ 
    (dashed and dotted lines for positive and negative values).   
%    The solid and dashed lines respectively correspond to 
%    positive and negative $v_R$. 
%    {\it top right:} Plasma $\beta (=p/p_{\rm B})$ value. 
%    {\it middle right:} Shakura-Sunyaev $\alpha$ value (solid) consisting 
%    of the Maxwell stress (dashed) and Reynolds stress (dotted) (see 
%    Equation \ref{eq:alpha}). 
    \label{fig:tave2}
  }
\end{figure}

\begin{figure}%[h]
  \begin{center}
    \hspace{0.5cm}\includegraphics[height=0.2\textheight]{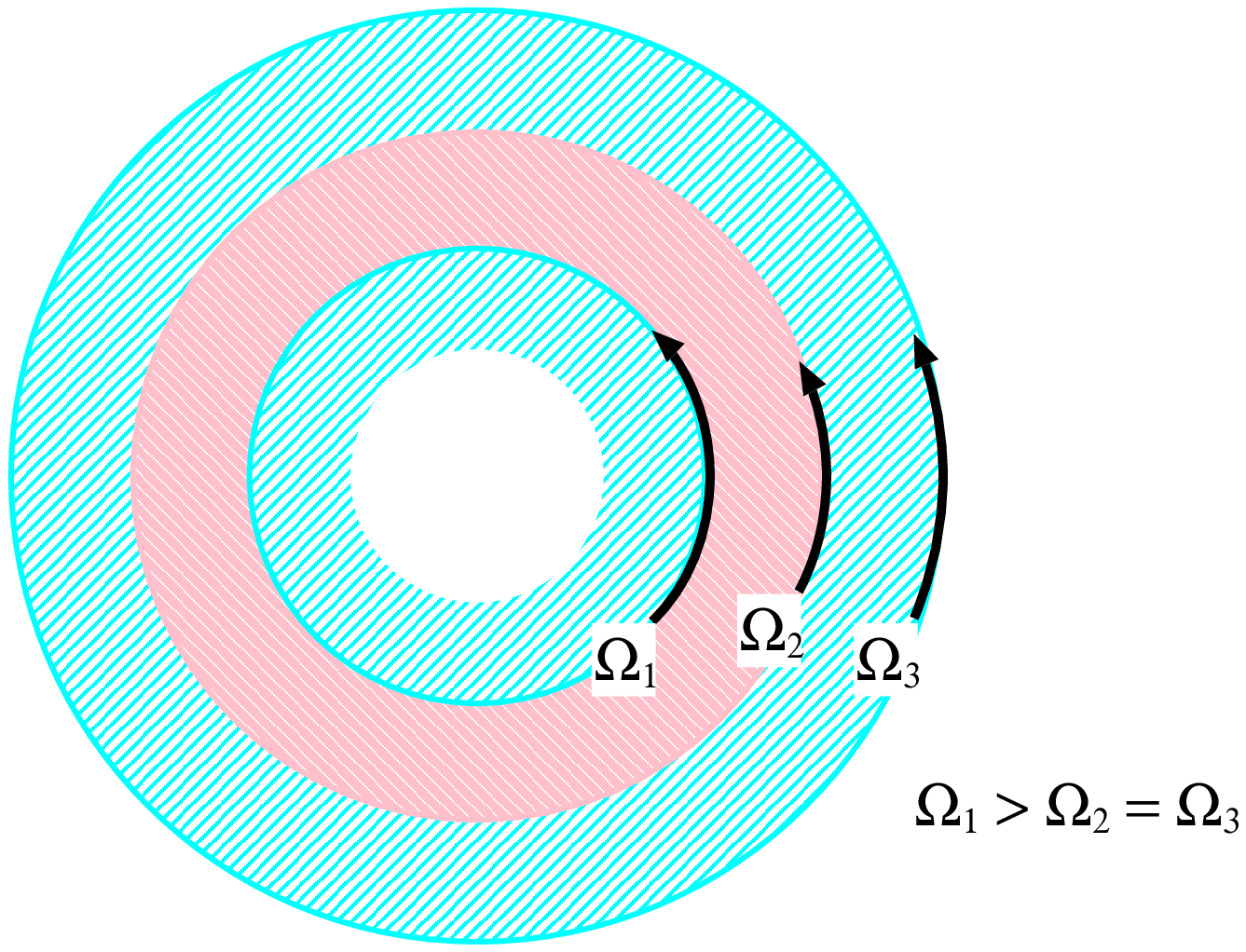}\\
    \vspace{0.1cm}
    \hspace{0.5cm}\includegraphics[height=0.2\textheight]{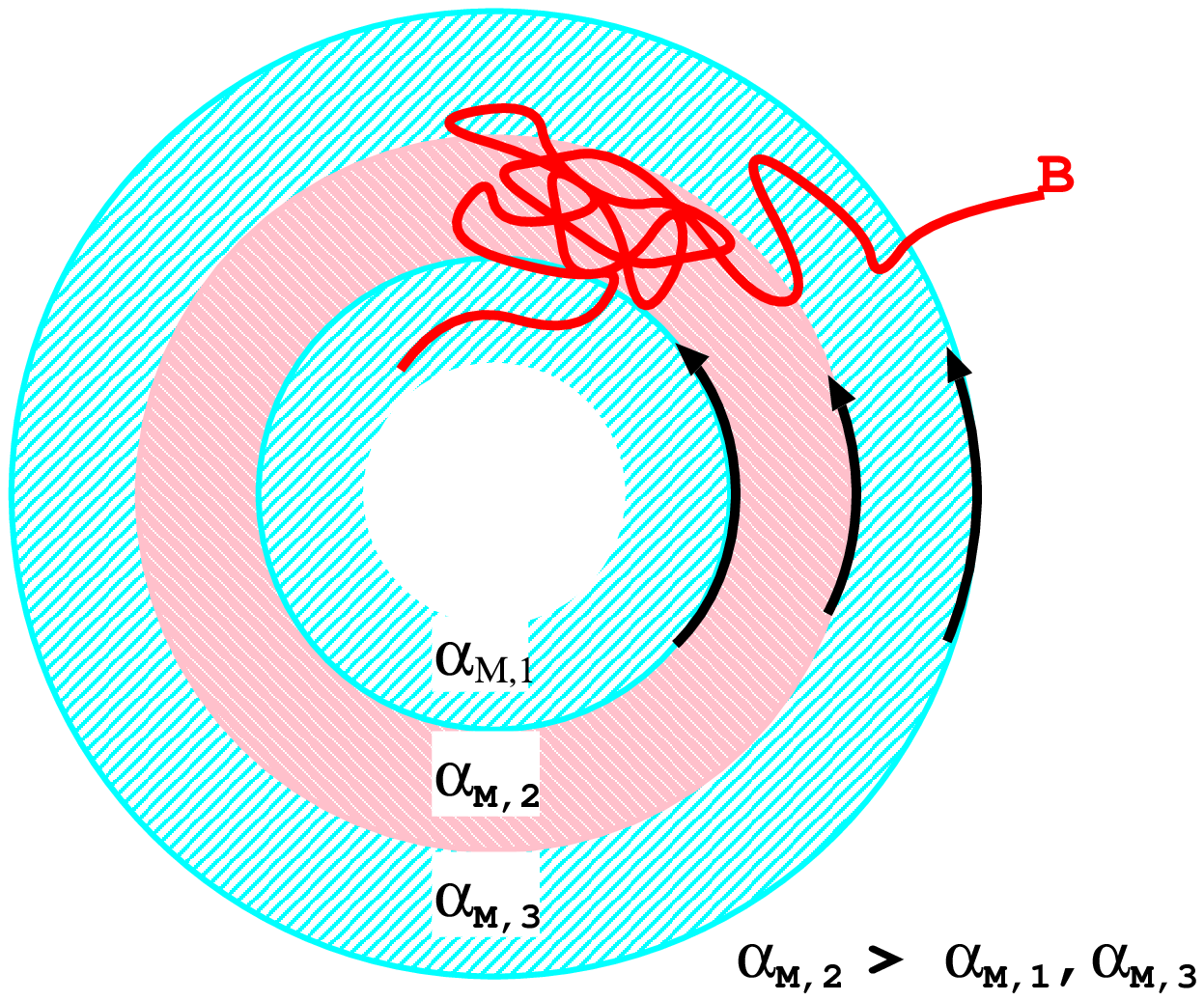}\\
    \vspace{0.1cm}
    \includegraphics[height=0.2\textheight]{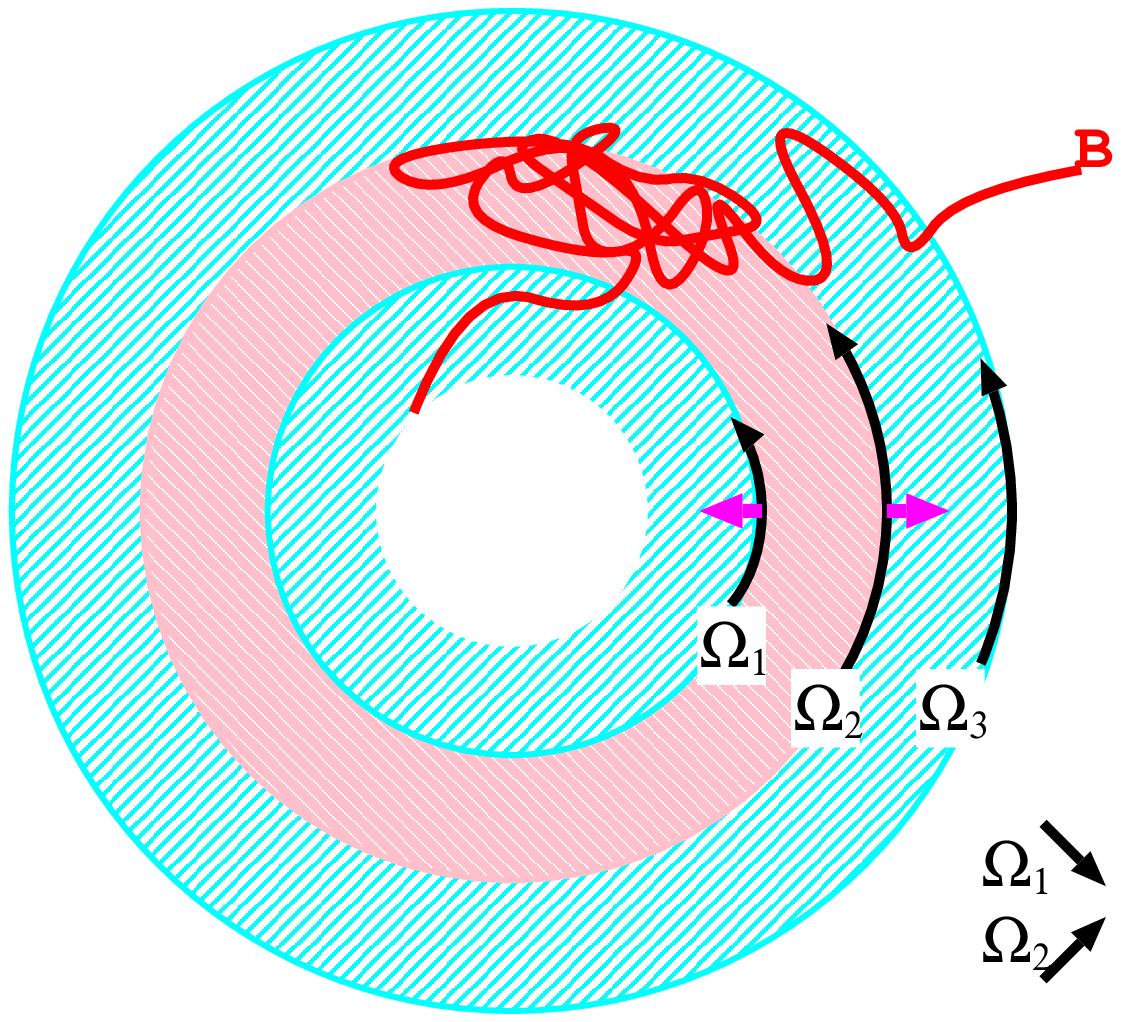}
  \end{center}
  \caption{Simplified schematic picture of the mechanism that drives radial 
    flows. 
    Let us suppose three rings, `1', `2', and `3', that rotate with rotation 
    frequency, $\Omega_1$, $\Omega_2$, and $\Omega_3$, at the outer edge of 
    each ring. 
    If $\Omega_1 > \Omega_2 = \Omega_3$, differential rotation is 
    more intense in the red region (Region 2). 
    The magnetic field is amplified more effectively there by the MRI and 
    the field-line stretching, and the Maxwell stress, $\alpha_{\rm M}$, 
    (Equation \ref{eq:alpha}) is larger than those in the neighboring zones, 
    which gives the faster outward transport of the angular momentum in 
    the red region. As a result, $\Omega_1$ decreases 
    and $\Omega_2$ increases. From the radial force balance, the inner edge 
    of the red region moves inward because of the decrease of the centrifugal 
    force, while the outer edge moves outward by the increase of the
    centrifugal force. Although the inhomogeneity of angular momentum 
    transport is distributed in a far more complicated fashion in our 
    simulation, the essential point can be explained by this simple picture. 
    \label{fig:cartoon1}
  }
\end{figure}

\begin{figure}%[h]
  \begin{center}
    \includegraphics[height=0.2\textheight]{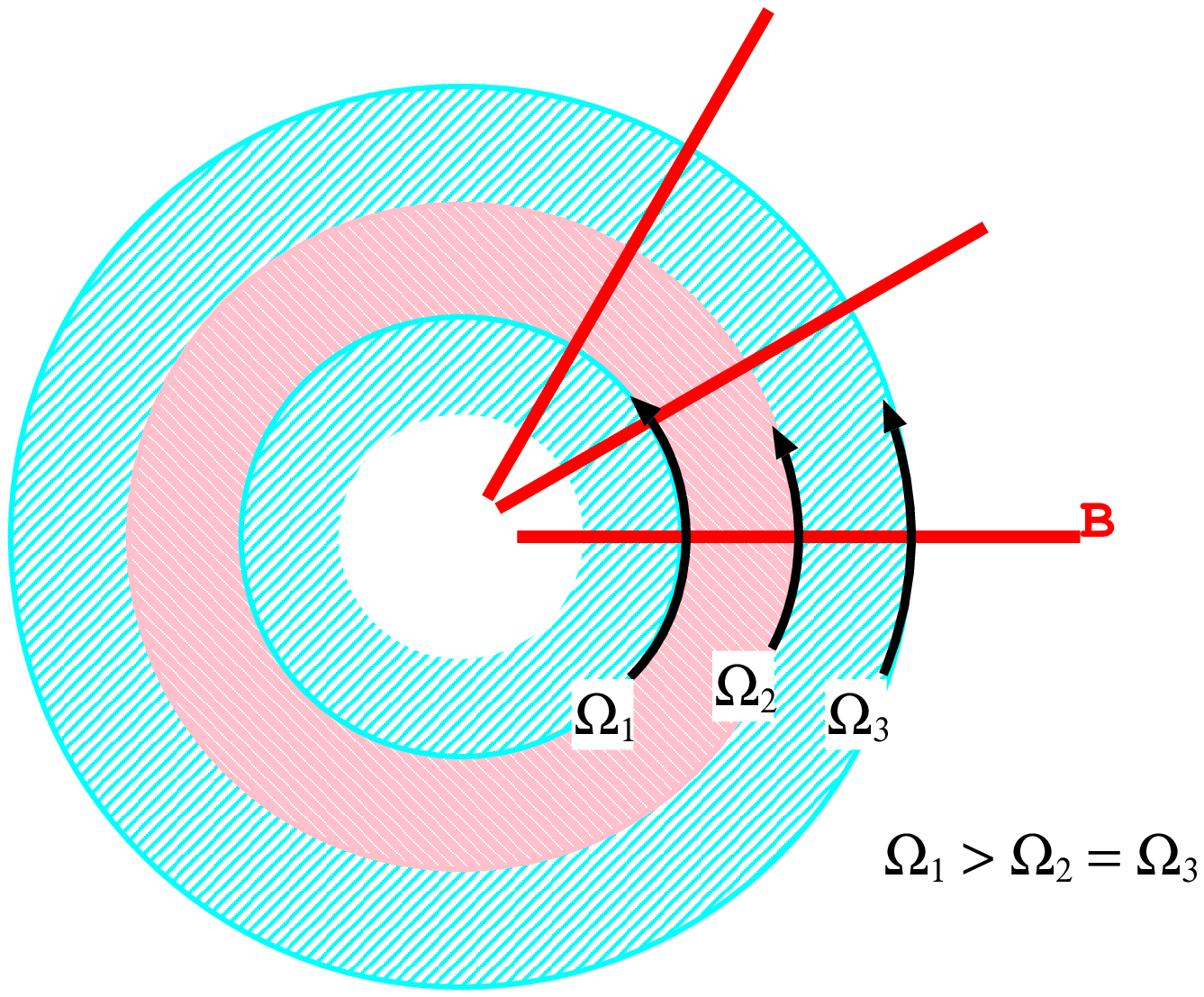}\\
    \includegraphics[height=0.22\textheight]{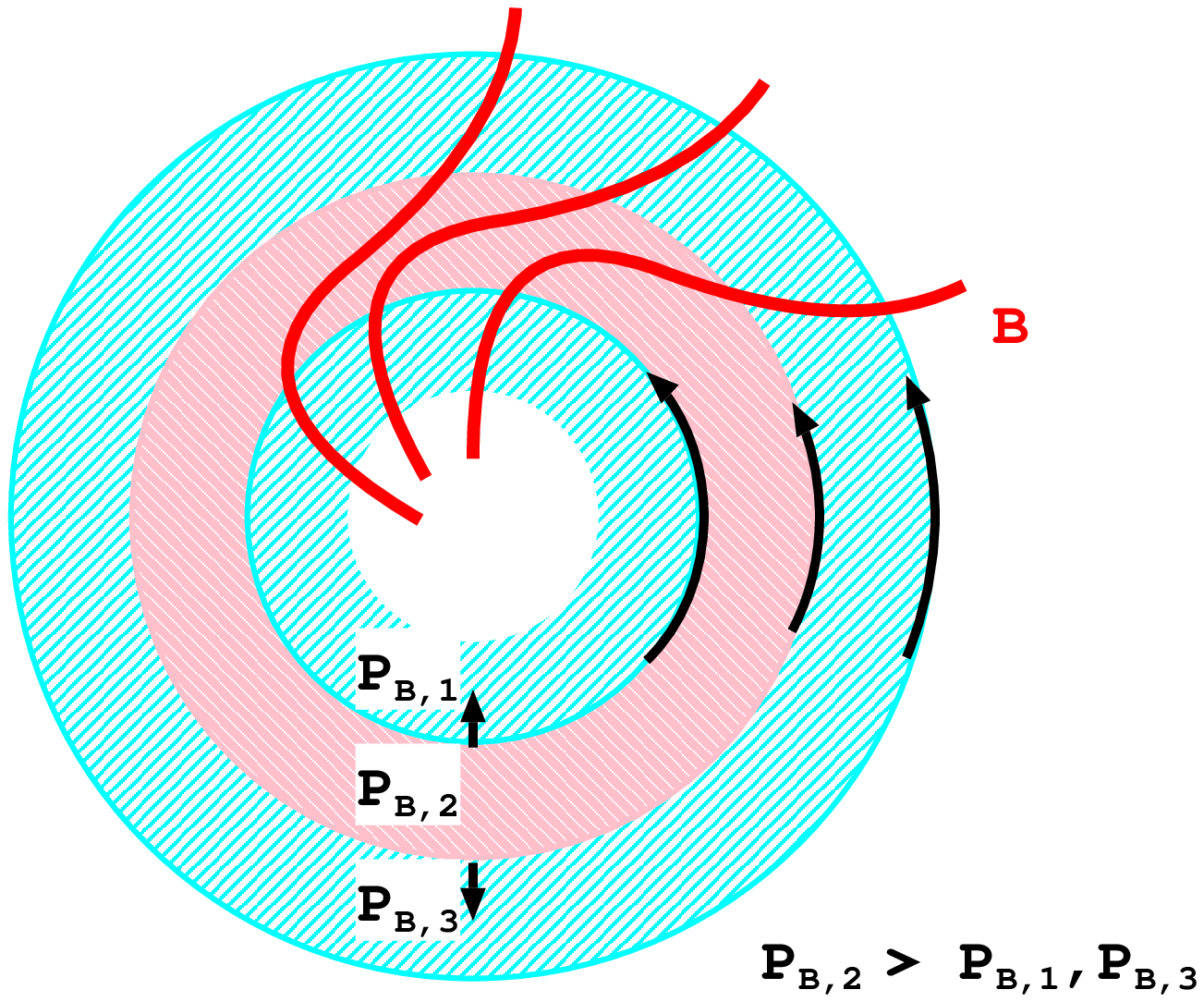}
  \end{center}
  \caption{Additional mechanism that drives radial flows. 
    The initial setting is the same as in Figure \ref{fig:cartoon1}. 
    Radial magnetic field, which is 
    triggered by MRI in our simulation, is amplified to general toroidal field 
    by the differential rotation selectively in the red region (Region 2). 
    The magnetic pressure, $p_{\rm B}$, of the toroidal field there
    is larger than $p_{\rm B}$ in the neighboring regions. 
    Then, radial flows are generated by the magnetic pressure gradient force.  
    \label{fig:cartoon2}
  }
\end{figure}
%\fi

We examine the time-dependency of the parallelogram-like feature in 
Figure \ref{fig:p-v_tdp}. The solid lines in the top panel shows the 
time-evolution of the maximum and minimum velocities ($v_{\rm max}$ \& 
$v_{\rm min}$) for column density $>0.5\times 10^{21}$cm$^{-2}$ 
%in the direction of the line of sight toward 
at $l=0$ degree; 
they correspond to the positions where the upper and lower sides of the 
parallelogram cross the vertical line of $l=0$ degree in the $l-v$ diagram.  
The figure shows that $v_{\rm max}$ and $v_{\rm min}$ vary within 
50-150 km s$^{-1}$ with time. Basically $v_{\rm max}$ and $v_{\rm min}$ change 
independently of each other, and then, the shape of the symmetric parallelogram 
is not always kept but distorted with time as seen in Figure \ref{fig:p-v1}. 

%For instance, the bottom right 
%panel of Figure \ref{fig:p-v1} shows a bump at $l\approx 1$ deg. This reflects 
%the radial motion of a non-axisymmetric dense region at $R\approx 0.1$ kpc, 
%which we will discuss, related to the central molecular zone, in 
%\S \ref{sec:CMZ}. 

In Figure \ref{fig:p-v_tdp}, the time evolution of the motion of the gas at 
$(x,y)=(0,\mp 0.5)$ kpc is also shown (dashed lines) for comparison. 
These lines indicate that the direction of the radial motion tends to be 
outward ($v<0$ for the front side and $v>0$ for the back side), 
whereas inward motions are also seen occasionally.  

The bottom panel presents the slope of the parallelogram measured near
$l=0$, which is derived from the average in the region with $-0.72 < l < 0.72$ 
deg, corresponding to $x \approx \pm 100$ pc. The solid and dashed lines 
denote the top and bottom sides of the parallelogram, which mostly reflect 
far and near sides with respect to the Galactic center, respectively. 
The top and bottom sides change independently because of the non-axisymmetric 
excitation of radial flows.  
The slopes vary from 15 to 90 km s$^{-1}$deg$^{-1}$, which covers the observed 
slope, $\approx 20$ km s$^{-1}$deg$^{-1}$, for the top side of the parallelogram 
and $\approx 55$ km s$^{-1}$deg$^{-1}$ for the bottom side \citep{tor10a}.

%\newpage
\subsection{Origin of Noncircular Motion \\--Roles of Magnetic Field--}     
\label{sec:oncm}

%We inspect properties of the MHD turbulence and its effect on the global 
%dynamics of the Galactic bulge and disk, particularly focusing on the 
%mechanism that drives the large fluctuation in $v_{R}$. 
Our MHD simulation shows that noncircular motions are excited, even though 
the axisymmetric gravitational potential is adopted. 
Here we discuss the roles of the magnetic field in driving the radial gas 
flows by inspecting the numerical data.
In Figures \ref{fig:tave1} and \ref{fig:tave2} we present the radial profile 
of various quantities averaged over azimuthal ($\phi$) and vertical ($z$) 
directions and over time $t$. We take the time-average of 
the final stage of the simulation from $t_1=27\pi t_{\rm unit}=436.68$ Myr 
to $t_2 =30\pi t_{\rm unit}= 485.21$ Myr. Although the magnetic field is  
already amplified to achieve the saturated state in the bulge region, it is 
still developing outside the bulge (Figure \ref{fig:tevol}). Therefore, 
in these figures we focus on the inner region, $R \lesssim 2$ kpc. 
%, because MHD turbulence is still developing in the outer region. 
We take the average of a physical quantity, $A$, as follows:
\begin{equation}
\langle A\rangle = \int_{t_1}^{t_2}dt\int_{-\pi}^{\pi}d\phi\int_{z_1}^{z_2}dz A
/[2\pi (t_2-t_1)(z_2-z_1)], 
\label{eq:ave1}
\end{equation}
where we set $z_1=-1$ and $z_2=+1$ kpc.
For the quantities concerning velocity, we adopt density-weighted averages; 
the average of flow velocity and root-mean squared (rms) velocity is taken by 
\begin{equation}
\langle v\rangle = \frac{\langle\rho v\rangle}{\langle \rho \rangle}, 
\label{eq:ave2}
\end{equation}
and
\begin{equation}
\sqrt{\langle v^2\rangle} = \sqrt{\frac{\langle\rho v^2\rangle}{\langle \rho 
\rangle}}, 
\label{eq:ave3}
\end{equation}
while the simple average is taken for the quantities on magnetic field. 
%Equation (\ref{eq:ave1}).

Figure \ref{fig:tave1} presents the plasma $\beta$ value, 
\begin{equation}
\langle \beta \rangle = \frac{\langle p \rangle}{\langle p_{\rm B}\rangle} 
=\frac{8\pi \langle p \rangle}{\langle B^2\rangle}
\end{equation}
which is the ratio of gas pressure to magnetic pressure (top panel), 
and the $\alpha$ value (bottom panel; solid line) 
\citep{ss73}, 
\begin{equation}
\langle \alpha \rangle = \langle \alpha_{\rm R} \rangle + \langle \alpha_{\rm M} 
\rangle = \frac{\langle\rho v_R \delta v_{\phi}\rangle}{\langle p\rangle} 
- \frac{\langle B_R B_{\phi}\rangle}{4\pi\langle p\rangle} 
\label{eq:alpha}
\end{equation}
which is the sum of Reynolds ($\alpha_{\rm R}$; dotted line) and Maxwell 
($\alpha_{\rm M}$; dashed line) stresses. 
The fluctuation component, $\delta v_{\phi}$, is derived by 
subtracting the mean rotation, 
\begin{equation}
%\sqrt{\langle\delta v_{\phi}^2\rangle} = \sqrt{\langle v_{\phi}^2\rangle
%-\langle v_{\phi}\rangle^2}, 
\delta v_{\phi} = v_{\phi} - \langle v_{\phi}\rangle . 
\label{eq:dvphi}
\end{equation}
%where all the $\langle \rangle$'s here are density-weighted averages, 
%Equations (\ref{eq:ave2}) and (\ref{eq:ave3}).

The top panel of Figure \ref{fig:tave1} shows that the plasma $\beta$ is 
below 100 in the bulge region, $R<1$ kpc, which is much smaller than the 
initial value, $\beta \approx 10^{3}-10^{5}$, as a consequence  of the 
amplification of the initial weak magnetic field. 
The bottom panel shows that a large $\alpha$ value $>10^{-2}$ is obtained. 
In the inner bulge, $R < 0.3$ kpc, the Maxwell stress dominates the Reynolds 
stress. In the outer bulge, the Reynolds stress exceeds the Maxwell stress, 
and they both show bumpy structures, which reflects the complicated rotation 
profile as discussed from now. 

Figure \ref{fig:tave2} (a) presents the time and $\phi$--averaged strength 
of the radial differential rotation, $\partial \ln \Omega / \partial \ln R$, 
at the Galactic plane.  Note that $\partial \ln \Omega / \partial \ln R=0, -1, -3/2$ 
correspond to rigid-body, flat, and Keplerian rotations, respectively. 
The MRI is active when $\partial \ln \Omega / \partial \ln R <0$, and 
its maximum growth rate to generate radial magnetic field is given by 
\begin{equation}
\frac{\omega_{\rm MRI,max}}{\Omega} = \frac{1}{2}\left|\frac{\partial \ln \Omega}
{\partial \ln R}\right|
\label{eq:mri}
\end{equation}  
\citep{bh91}. 
The amplification of toroidal field by winding in the radial differential 
rotation follows \citep[e.g.,][]{bh98} 
\begin{equation}
\frac{1}{\Omega}\frac{\partial B_{\phi}}{\partial t} 
= B_R \frac{\partial \ln \Omega}{\partial \ln R}.
\label{eq:winding}
\end{equation}
Equations (\ref{eq:mri}) and (\ref{eq:winding}) show that stronger differential 
rotation amplifies magnetic field faster by both MRI and field-line 
stretching.

Figure \ref{fig:tave2}(a) shows that the initial rotation profile is more or 
less kept at the later time. The local minimum of $\partial \ln \Omega 
/ \partial \ln R$ is located at $R\approx 0.6$ kpc, where the rotation 
rapidly decreases with $R$ because of the contribution from the gas 
pressure-gradient force (\S \ref{sec:setup}).  
In this region, the magnetic field is effectively amplified on account of 
the strong differential rotation. The magnetic field strength is amplified to 
$|B|=\sqrt{B_R^2 + B_{\phi}^2 + B_z^2}\approx 0.1-1$ mG 
in the bulge region, which
is consistent with an observational lower limit, $>50$ $\mu$G \citep{cro10} and 
an empirical estimate, $\sim$ mG, based on multiple observations \citep{fer09}.

The effective amplification of the magnetic field around 
the location for the maximum differential rotation at $R\approx 0.6$ kpc 
gives a bump in the Maxwell stress, $\alpha_{\rm M}$ (Equation \ref{eq:alpha}) 
as shown by the dashed line in the bottom panel of Figure \ref{fig:tave1}. 
$\alpha_M$ determines the outward transport of angular momentum 
\citep[e.g.,][]{lp74} by magnetic field. The inhomogeneous distribution of 
$\alpha_{\rm M}$ indicates that spatially dependent gain or loss of the angular 
momentum takes place, which is illustrated by a simplified cartoon in Figure 
\ref{fig:cartoon1}; in a region with the gain of angular momentum the rotation 
is accelerated, and vice versa. This leads directly to the increase or decrease 
of the centrifugal force, which excites radial motion. Although what is taking 
place in our simulation is more complicated, namely $\alpha_{\rm M}$ is 
distributed in a non-axisymmetric manner, and both outward and inward flows 
are generated even at the same $R$ as exhibited in Figure \ref{fig:faceon1}, 
this simple conceptual cartoon in Figure \ref{fig:cartoon1} explains the 
essential point.

In addition to the inhomogeneous transport of angular momentum, magnetic 
pressure also contributes to the excitation of radial flows. 
In the outer bulge region near $R\approx 1$ kpc, the differential rotation is 
weak and the amplification of magnetic field is suppressed there. 
Figure \ref{fig:tave2}(b) exhibits the rapid decrease of the magnetic field 
strength with $R$ in $R\gtrsim 0.5$ kpc.
%both the toroidal ($\phi$), which is mainly by the winding 
%(Equation \ref{eq:winding}), and the radial ($R$) component, owing to the 
%MRI (Equation \ref{eq:mri}), of the magnetic field there.
The rapid decrease of $B_{\phi}^2$ causes the radial pressure gradient, 
$-\frac{1}{8\pi\rho}\frac{\partial B_{\phi}^2}{\partial R}$, which drives 
outward radial flows (Figure \ref{fig:cartoon2}).  

We have discussed the two types of the processes, 
the inhomogeneous angular momentum transport and the magnetic pressure-gradient 
force, that generate radial flows. By inspecting each term in the momentum 
equation (\ref{eq:mom}) numerically, we found that the former dominates 
the latter by $\approx 2:1$. 

Figure \ref{fig:tave2}(c) shows that the direction of the mean flow is outward 
(dashed line $\langle v_R \rangle >0$) in $0.3\lesssim R \lesssim 2$ kpc,  
which is expected from both processes. This radially outward 
flow also transports the angular momentum, which is a reason in part why the 
Reynolds stress (the bottom panel of Figure \ref{fig:tave1}) shows a bump 
in $0.5\lesssim R \lesssim 1$ kpc.
Compared to the mean flow, the rms velocity, 
$\sqrt{\langle v_R^2\rangle}$, (solid line) at $R\approx 0.5$ kpc 
is quite large, $\approx 40$ km s$^{-1}$, because the MRI-triggered 
turbulence drives both inward and outward $v_R$ intermittently \citep[]
[see also movies of Figures \ref{fig:snp1} and \ref{fig:faceon1}]{sano01}, 
as discussed so far. 
This is the main reason why the simulated $l$--$v$ diagrams show a thick 
parallelogram-like shape (Figure \ref{fig:p-v1}).

\subsection{Non-axisymmetric Structure}

%\if0
\begin{figure}%[h]
  \vspace{-1cm}
%  \begin{center}
    \hspace{-2cm}
    \includegraphics[height=0.36\textheight]{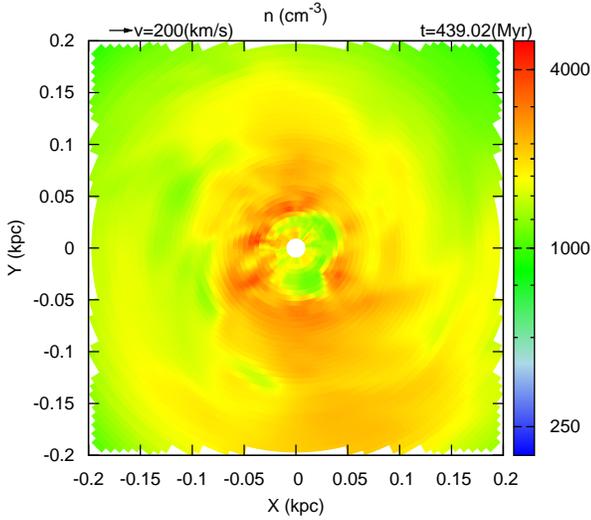}\\
%    \includegraphics[height=0.36\textheight]{../worksphmpi/glbdsk_velfld28inv_t27-10pi_hr1.eps}\\
%  \end{center}
    \vspace{-1cm}
  \caption{
    Density structure in the Galactic center region at $t=439.02$ Myr. 
    (Zoomed-in view of the left panel of Figure \ref{fig:snp1}.)
    \label{fig:cmz}
  }
\end{figure}
%\fi
As we have discussed in the previous subsection, the velocity field shows 
non-axisymmetric radial flows because of the intermittency of the MRI-triggered 
turbulence. 
The density distribution also shows non-axisymmetric inhomogeneity, which is 
typical for the MRI turbulence as well \citep{si14}, in the 
central region as exhibited in Figure \ref{fig:cmz}.
% that presents the snapshot density at the midplane at $t=439.02$ Myr. 
One can see multiple arm-like structure, and when measured at
$R=50$ pc, the density varies largely within, $1500 < n < 3800$ cm$^{-3}$, 
which roughly corresponds to $750 < n_{\rm H_2} < 1900$ cm$^{-3}$ in molecular 
number density. 

It is well known that the Milky way possesses the central molecular zone (CMZ), 
which contains molecular gas with mass $\approx (5-10)\times 10^7 M_{\odot}$ 
and occupies a volumetric filling factor $\gtrsim 0.1$ within 
$R<200$ pc \citep{ms96}.
The CMZ consists of non-axisymmetric clouds with complicated structure
\citep{oka05,kru15}. 
Our simulation implies that such non-axisymmetric clouds are a natural outcome 
of magnetic activity in the Galactic center.
The molecular number density estimated from our simulation above is smaller 
than typical observed values, $\sim 10^4$ cm$^{-3}$ \citep[e.g.,][]{nag07}.
This is because our simulation assumes the one-fluid gas with the locally 
isothermal equation of state. 
In $R<200$ pc, the temperature is assumed to be constant $\sim 10^5$ K in the 
simulation, which is much higher than the observed typical temperature, 
30--200 K, of molecular clouds in the CMZ \citep[e.g.][]{ms96}. 
We need to treat the energy equation with heating and cooling to deal with 
different phases of the gas material such as molecular clouds 
and warm neutral media that respectively possess different filling factors. 
The temperature of dense regions in the simulation 
would be much lower if the radiative cooling was taken into account, 
and accordingly the density would be higher to satisfy the pressure balance. 
%If we took into account the heating and cooling, 
%the density contrast would be much higher reflecting temperature difference; 
%the density of molecular clouds would be higher than the obtained maximum, 
%$N_{\rm H}=??$ in our simulation. 

\subsection{Outflows} 

%\if0
\begin{figure}%[h]
  \vspace{-1cm}
  \begin{center}
    \includegraphics[height=0.38\textheight]{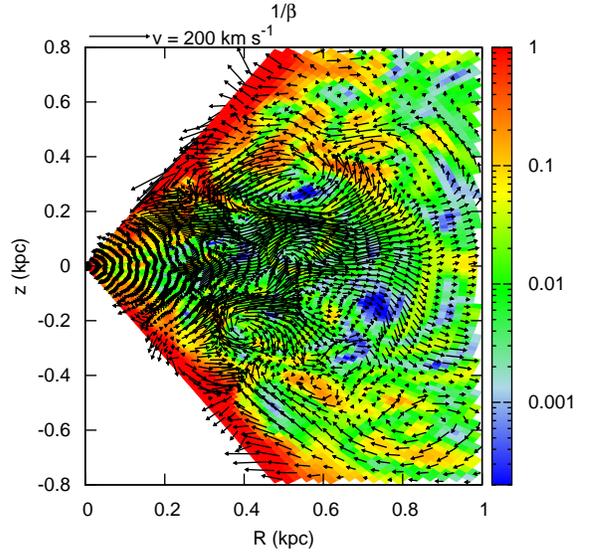}
  \end{center}
  \vspace{-1cm}
  \caption{Vertical cut at $\phi=\pi/2$ of $1/\beta$ (colours) and velocity 
    field (arrow) at $t=439.02$ Myr.
\label{fig:edgeon}
}
\end{figure}
%\fi

Figure \ref{fig:edgeon} presents the snapshot of a vertical cut 
at $t=439.02$ Myr, Here the colours indicate $1/\beta = p_{\rm B}/p$; 
redder colours correspond to regions with relatively strong magnetic pressure 
compared to gas pressure. 
The figure shows that flow patterns are quite complicated; 
one can see both inward flows to the bulge and outward streams from it. 
Although both inward and outward streams coexist, if we integrate the mass 
flux in space and time, the direction of the net gas flow is outward 
from the bulge; for 
example the mass loss rate measured at the spherical surface, $r=1.0$ kpc, 
averaged during $t=436.68 - 485.21$ Myr, gives $\dot{M} = \int dt \int d\phi 
\int d\theta \int dr \rho v_r r^2 \sin \theta \approx$ 70 $M_{\odot}$ yr$^{-1}$, 
and the kinetic energy luminosity, $\approx \frac{1}{2}\dot{M} v_r^2 
\sim 10^{38}$ erg s$^{-1}$.
It should be noted that this outflow rate might be suppressed 
when a stellar-bar potential was taken into account \citep{cha00}. 

In addition to radial flows, upward flows (winds) stream out of 
the $\theta$ surfaces of the simulation box.  
The typical speed of these flows is several tens km s$^{-1}$, and in some 
regions the speed exceeds 100 km s$^{-1}$. 
%Such vertical outflows possibly explain the observed vertical 
%structure of molecular gas associated with the DHN \citep{eno14,tor14b}.
Such vertical outflows possibly explain the observed vertical structure of molecular gas associated with the DHN \citep{eno14,tor14b}. Recent observation further reveals CO emission at high Galactic latitude, $b$, including several filamentary features in addition to diffuse extended halo-like CO gas up to 2$^\circ$ in $b$ above and below the CMZ \citep{tor14a}. 
A plausible interpretation is that these high-$b$ features are also driven by the Parker instability; although they are filamentary 
without a clear loop-like shape having two foot points, the MHD numerical simulations by \citet{mac09} show that magnetic-flotation loops are generally not symmetric and can be seen as an open single filament depending on the ambient density/field distribution and evolutionary effect. 
Further signature is associated with the double helix nebula (DHN, hereafter) toward $(l,b)=(0.0\; {\rm deg.}, 0.8\; {\rm deg.})$ \citep{mor06}, where a column of molecular gas of 200-pc height is found to be associated with the DHN which is nearly vertical to the plane \citep{eno14,tor14b}. Such a feature is explained as created by a magnetic tower formed above and below the central black hole according to preceding numerical simulations \citep{kat04,mac09}.

Recently, the existence of outflows or winds in the Galactic center region 
is extensively discussed \citep{eve08,cro11}, and it might be closely linked 
to the formation mechanisms of the Fermi bubbles via magnetic activity 
\citep{car13}. 
Our simulation is not directly applicable to the Fermi bubbles 
since the simulation does not cover the region near the polar axis. 
However, the above rough estimate of the energetics of the outflow in our 
simulation shows that magnetic activity is a reliable mechanism in driving 
outflows from the bulge. 

\section{Summary and Discussion}
We have investigated the excitation of radial gas flows in the Galactic center 
region by the 3D global MHD simulation with the axisymmetric gravitational 
potential. %by \citet{mn75}. 
The initial weak vertical magnetic field is amplified by the MRI at the 
beginning, and in addition eventually by the Parker instability and the 
field-line stretching owing to the differential rotation. In the bulge region, 
the final field strength is $\gtrsim 0.5$ mG, and the plasma $\beta \sim 1-40$. 
Because of the gas pressure-gradient force, the equilibrium rotation 
frequency is non-monotonic with radial distance. The amplification of 
the magnetic field is systematically more effective in regions with stronger 
differential rotation. 
Then, the transport of the angular momentum is spatially dependent, and 
%the rotation speed is accelerated in some regions and decelerated in other 
%regions inhomogeneously. 
the acceleration or deceleration of the rotation speed occurs depending on 
regions. This breaks the radial pressure balance owing to the change of the 
centrifugal force, and excites radial flows of the gas in a time-dependent and 
non-axisymmetric manner. 
In addition, the radial component of the magnetic pressure 
gradient is produced, which also excites radial flows.  
As a result, the simulated $l$--$v$ diagram exhibits a time-dependent and 
asymmetric parallelogram shape. 
The rotation curve near the Galactic center exhibits 
complicated features \citep{sof13}. Our simulation implies that stochastically 
excited radial motion might be contaminated in these features. 

In order to interpret the parallelogram in the CMZ, the elliptical orbit in 
the stellar-bar potential has been used as the only viable model \citep{bin91}. 
The model may not be fully appropriate as the interpretation from an 
observational point. First, the bar potential alone is not able to create 
high-$b$ features up to 200 pc above the plane; these “loops” are 
better explained by the Parker instability \citep{fuk06}. 
There is no other driving mechanism known to expel gas up to that height by only 
stellar gravity. Second, the parallelogram is not symmetric in the $l$-$v$ 
diagram; the velocity gradient in the positive velocity, $\sim$20 km s$^{-1}$ 
deg$^{-1}$, is significantly different by a factor of 3 from that in the 
negative velocity, $\sim$55 km s$^{-1}$ deg$^{-1}$ \citep{tor10a}. 
The present model shows that the velocity gradient of the parallelogram 
is naturally produced as a time dependent manner, whereas it is not clear 
how such a difference is explained in the bar potential model. 

On the other hand, we see some features that are not reproduced in the present 
simulation. The large broad features like the clump 2 and the 5.5 deg feature 
are not reproduced in the present simulation.
In our simulation we have assumed one-fluid gas and neglected 
heating and cooling with a simplified treatment of the locally isothermal gas. 
%, namely we have assumed spatially dependent but time-independent temperature. 
In order to treat dense molecular gas in a quantitative fashion, we should 
replace these simplifications. 
If radiative cooling, as well as adiabatic heating and cooling, is included, 
large density contrast between cool molecular gas and warm/hot gas will be 
naturally reproduced, which may explain the observed fine-scale features.

%it is also necessary to solve an energy equation with radiative cooling 
%in addition to adiabatic heating and cooling. 

We have shown that the magnetic activity possibly produces 
the overall trend of the observed asymmetric parallelogram-shape in the $l-v$ 
diagram even though the axisymmetric gravitational potential is considered.  
%In our simulation, we dare to adopt the axisym metric potential so as to 
%concentrate on the effect of the magnetic field on noncircular motion of 
%the gas. 
%For more realistic simulations to explain these observed features in the 
%$l$--$v$ diagram in detail, it would be desirable to take into account 
%a bar potential in the MHD simulation \citep[see][for two dimensional 
%simulations]{ks12}, and compare the effect of the bar to that of the 
%magnetic field.
It is also desirable to test quantitatively the importance of the magnetic 
activity in contrast to role of the stellar bar potential in the noncircular 
motion of the gas after the cooling/heating effect is taken into account 
in our simulation. 

Although in this paper we have focused on the bulge region of our Milky Way, 
results of our simulation can be applied to other galaxies. 
For example, \citet{sak06} observed NGC 253 by the Submillimeter Array and 
reported that they found an expanding circumnuclear disk with 
$\sim 50$ km s$^{-1}$. The obtained position velocity diagrams show asymmetric 
parallelogram features in different wavelengths. These features are 
qualitatively similar to but quantitatively different from those obtained 
in the Milky Way. 
%A possible explanation is that the difference reflects the 
%detailed profiles of the rotation frequencies based on our simulation.  
%Based on our simulation, the detailed profile of the position--velocity 
%diagram changes with time. 
A possible explanation of the difference is the time variability inhering in 
the MHD turbulence, based on our simulation. 

\section*{Acknowledgments}
This work was supported by Grants-in-Aid for Scientific Research 
from the MEXT of Japan, 24224005 (PI: YF). 
Numerical simulations in this work were carried out at the Cray XC30 
(ATERUI) operated in CfCA, National Astrophysical Observatory of Japan, 
and the Yukawa Institute Computer Facility, SR16000. 
The authors thank the anonymous referee for constructive comments to improve 
the paper.
TKS thanks Prof. Shu-ichiro Inutsuka and Dr. Kazunari Iwasaki 
for fruitful discussion.

\pageref{lastpage}

%\end{CJK*}

\begin{thebibliography}{84}
\expandafter\ifx\csname natexlab\endcsname\relax\def\natexlab#1{#1}\fi

\bibitem[{{Baba}, {Saitoh} \& {Wada}(2010){Baba}, {Saitoh}, \& {Wada}}]{bab10}
{Baba} J., {Saitoh} T.~R., {Wada} K., 2010, \pasj, 62, 1413

\bibitem[{{Balbus} \& {Hawley}(1991)}]{bh91}
{Balbus} S.~A., {Hawley} J.~F., 1991, \apj, 376, 214

\bibitem[{{Balbus} \& {Hawley}(1998)}]{bh98}
{Balbus} S.~A., {Hawley} J.~F., 1998, Reviews of Modern Physics, 70, 1

\bibitem[{{Bally} {et~al}\mbox{.}(1987){Bally}, {Stark}, {Wilson}, \&
  {Henkel}}]{bal87}
{Bally} J., {Stark} A.~A., {Wilson} R.~W., {Henkel} C., 1987, \apjs, 65, 13

\bibitem[{{Bania}(1977)}]{ban77}
{Bania} T.~M., 1977, \apj, 216, 381

\bibitem[{{Binney} {et~al}\mbox{.}(1991){Binney}, {Gerhard}, {Stark}, {Bally},
  \& {Uchida}}]{bin91}
{Binney} J., {Gerhard} O.~E., {Stark} A.~A., {Bally} J., {Uchida} K.~I., 1991,
  \mnras, 252, 210

\bibitem[{{Blitz}(1994)}]{bli94}
{Blitz} L., 1994, in Astronomical Society of the Pacific Conference Series,
  Vol.~66, Physics of the Gaseous and Stellar Disks of the Galaxy, {King}
  I.~R., ed., p.~1

\bibitem[{{Blitz} \& {Spergel}(1991)}]{bl91}
{Blitz} L., {Spergel} D.~N., 1991, \apj, 379, 631

\bibitem[{{Bovy} {et~al}\mbox{.}(2012){Bovy}, {Allende Prieto}, {Beers},
  {Bizyaev}, {da Costa}, {Cunha}, {Ebelke}, {Eisenstein}, {Frinchaboy},
  {Garc{\'{\i}}a P{\'e}rez}, {Girardi}, {Hearty}, {Hogg}, {Holtzman}, {Maia},
  {Majewski}, {Malanushenko}, {Malanushenko}, {M{\'e}sz{\'a}ros}, {Nidever},
  {O'Connell}, {O'Donnell}, {Oravetz}, {Pan}, {Rocha-Pinto}, {Schiavon},
  {Schneider}, {Schultheis}, {Skrutskie}, {Smith}, {Weinberg}, {Wilson}, \&
  {Zasowski}}]{bov12}
{Bovy} J. {et~al.}, 2012, \apj, 759, 131

\bibitem[{{Carretti} {et~al}\mbox{.}(2013){Carretti}, {Crocker},
  {Staveley-Smith}, {Haverkorn}, {Purcell}, {Gaensler}, {Bernardi}, {Kesteven},
  \& {Poppi}}]{car13}
{Carretti} E. {et~al.}, 2013, \nat, 493, 66

\bibitem[{{Chandran}, {Cowley} \& {Morris}(2000){Chandran}, {Cowley}, \&
  {Morris}}]{cha00}
{Chandran} B.~D.~G., {Cowley} S.~C., {Morris} M., 2000, \apj, 528, 723

\bibitem[{{Chandrasekhar}(1961)}]{cha61}
{Chandrasekhar} S., 1961, {Hydrodynamic and hydromagnetic stability}. Oxford:
  Clarendon

\bibitem[{{Chuss} {et~al}\mbox{.}(2003){Chuss}, {Davidson}, {Dotson}, {Dowell},
  {Hildebrand}, {Novak}, \& {Vaillancourt}}]{chu03}
{Chuss} D.~T., {Davidson} J.~A., {Dotson} J.~L., {Dowell} C.~D., {Hildebrand}
  R.~H., {Novak} G., {Vaillancourt} J.~E., 2003, \apj, 599, 1116

\bibitem[{{Clarke}(1996)}]{cl96}
{Clarke} D.~A., 1996, \apj, 457, 291

\bibitem[{{Crocker} {et~al}\mbox{.}(2011){Crocker}, {Jones}, {Aharonian},
  {Law}, {Melia}, \& {Ott}}]{cro11}
{Crocker} R.~M., {Jones} D.~I., {Aharonian} F., {Law} C.~J., {Melia} F., {Ott}
  J., 2011, \mnras, 411, L11

\bibitem[{{Crocker} {et~al}\mbox{.}(2010){Crocker}, {Jones}, {Melia}, {Ott}, \&
  {Protheroe}}]{cro10}
{Crocker} R.~M., {Jones} D.~I., {Melia} F., {Ott} J., {Protheroe} R.~J., 2010,
  \nat, 463, 65

\bibitem[{{Dame}, {Hartmann} \& {Thaddeus}(2001){Dame}, {Hartmann}, \&
  {Thaddeus}}]{dam01}
{Dame} T.~M., {Hartmann} D., {Thaddeus} P., 2001, \apj, 547, 792

\bibitem[{{de Vaucouleurs}(1964)}]{dev64}
{de Vaucouleurs} G., 1964, in IAU Symposium, Vol.~20, The Galaxy and the
  Magellanic Clouds, {Kerr} F.~J., ed., p. 195

\bibitem[{{Enokiya} {et~al}\mbox{.}(2014){Enokiya}, {Torii}, {Schultheis},
  {Asahina}, {Matsumoto}, {Furuhashi}, {Nakamura}, {Dobashi}, {Yoshiike},
  {Sato}, {Furukawa}, {Moribe}, {Ohama}, {Sano}, {Okamoto}, {Mori}, {Hanaoka},
  {Nishimura}, {Hayakawa}, {Okuda}, {Yamamoto}, {Kawamura}, {Mizuno}, {Onishi},
  {Morris}, \& {Fukui}}]{eno14}
{Enokiya} R. {et~al.}, 2014, \apj, 780, 72

\bibitem[{{Evans} \& {Hawley}(1988)}]{eh88}
{Evans} C.~R., {Hawley} J.~F., 1988, \apj, 332, 659

\bibitem[{{Everett} {et~al}\mbox{.}(2008){Everett}, {Zweibel}, {Benjamin},
  {McCammon}, {Rocks}, \& {Gallagher}}]{eve08}
{Everett} J.~E., {Zweibel} E.~G., {Benjamin} R.~A., {McCammon} D., {Rocks} L.,
  {Gallagher}, III J.~S., 2008, \apj, 674, 258

\bibitem[{{Feldmeier} {et~al}\mbox{.}(2014){Feldmeier}, {Neumayer}, {Seth},
  {Sch{\"o}del}, {L{\"u}tzgendorf}, {de Zeeuw}, {Kissler-Patig}, {Nishiyama},
  \& {Walcher}}]{fel14}
{Feldmeier} A. {et~al.}, 2014, \aap, 570, A2

\bibitem[{{Ferri{\`e}re}(2009)}]{fer09}
{Ferri{\`e}re} K., 2009, \aap, 505, 1183

\bibitem[{{Fromang} {et~al}\mbox{.}(2013){Fromang}, {Latter}, {Lesur}, \&
  {Ogilvie}}]{fro13}
{Fromang} S., {Latter} H., {Lesur} G., {Ogilvie} G.~I., 2013, \aap, 552, A71

\bibitem[{{Fujishita} {et~al}\mbox{.}(2009){Fujishita}, {Torii}, {Kudo},
  {Kawase}, {Yamamoto}, {Kawamura}, {Mizuno}, {Onishi}, {Mizuno}, {Machida},
  {Takahashi}, {Nozawa}, {Matsumoto}, \& {Fukui}}]{fuj09}
{Fujishita} M. {et~al.}, 2009, \pasj, 61, 1039

\bibitem[{{Fukui} {et~al}\mbox{.}(2006){Fukui}, {Yamamoto}, {Fujishita},
  {Kudo}, {Torii}, {Nozawa}, {Takahashi}, {Matsumoto}, {Machida}, {Kawamura},
  {Yonekura}, {Mizuno}, {Onishi}, \& {Mizuno}}]{fuk06}
{Fukui} Y. {et~al.}, 2006, Science, 314, 106

\bibitem[{{Genzel}, {Eisenhauer} \& {Gillessen}(2010){Genzel}, {Eisenhauer}, \&
  {Gillessen}}]{gen10}
{Genzel} R., {Eisenhauer} F., {Gillessen} S., 2010, Reviews of Modern Physics,
  82, 3121

\bibitem[{{Hayakawa} {et~al}\mbox{.}(1981){Hayakawa}, {Matsumoto}, {Murakami},
  {Uyama}, {Thomas}, \& {Yamagami}}]{hay81}
{Hayakawa} S., {Matsumoto} T., {Murakami} H., {Uyama} K., {Thomas} J.~A.,
  {Yamagami} T., 1981, \aap, 100, 116

\bibitem[{{Honma} {et~al}\mbox{.}(2012){Honma}, {Nagayama}, {Ando},
  {Bushimata}, {Choi}, {Handa}, {Hirota}, {Imai}, {Jike}, {Kim}, {Kameya},
  {Kawaguchi}, {Kobayashi}, {Kurayama}, {Kuji}, {Matsumoto}, {Manabe},
  {Miyaji}, {Motogi}, {Nakagawa}, {Nakanishi}, {Niinuma}, {Oh}, {Omodaka},
  {Oyama}, {Sakai}, {Sato}, {Sato}, {Shibata}, {Shiozaki}, {Sunada}, {Tamura},
  {Ueno}, \& {Yamauchi}}]{hon12}
{Honma} M. {et~al.}, 2012, \pasj, 64, 136

\bibitem[{{Kato}, {Mineshige} \& {Shibata}(2004){Kato}, {Mineshige}, \&
  {Shibata}}]{kat04}
{Kato} Y., {Mineshige} S., {Shibata} K., 2004, \apj, 605, 307

\bibitem[{{Kent}(1992)}]{ken92}
{Kent} S.~M., 1992, \apj, 387, 181

\bibitem[{{Kim} \& {Stone}(2012)}]{ks12}
{Kim} W.-T., {Stone} J.~M., 2012, \apj, 751, 124

\bibitem[{{Koda} \& {Wada}(2002)}]{kw02}
{Koda} J., {Wada} K., 2002, \aap, 396, 867

\bibitem[{{Kruijssen}, {Dale} \& {Longmore}(2015){Kruijssen}, {Dale}, \&
  {Longmore}}]{kru15}
{Kruijssen} J.~M.~D., {Dale} J.~E., {Longmore} S.~N., 2015, \mnras, 447, 1059

\bibitem[{{Kudo} {et~al}\mbox{.}(2011){Kudo}, {Torii}, {Machida}, {Davis},
  {Tsutsumi}, {Fujishita}, {Moribe}, {Yamamoto}, {Okuda}, {Kawamura}, {Mizuno},
  {Onishi}, {Maezawa}, {Mizuno}, {Tanaka}, {Yamaguchi}, {Ezawa}, {Takahashi},
  {Nozawa}, {Matsumoto}, \& {Fukui}}]{kud11}
{Kudo} N. {et~al.}, 2011, \pasj, 63, 171

\bibitem[{{Liszt} \& {Burton}(1978)}]{lb78}
{Liszt} H.~S., {Burton} W.~B., 1978, \apj, 226, 790

\bibitem[{{Liszt} \& {Burton}(1980)}]{lb80}
{Liszt} H.~S., {Burton} W.~B., 1980, \apj, 236, 779

\bibitem[{{Lynden-Bell} \& {Pringle}(1974)}]{lp74}
{Lynden-Bell} D., {Pringle} J.~E., 1974, \mnras, 168, 603

\bibitem[{{Machida} {et~al}\mbox{.}(2009){Machida}, {Matsumoto}, {Nozawak},
  {Takahashi}, {Fukui}, {Kudo}, {Torii}, {Yamamoto}, {Fujishita}, \&
  {Tomisaki}}]{mac09}
{Machida} M. {et~al.}, 2009, \pasj, 61, 411

\bibitem[{{Machida} {et~al}\mbox{.}(2013){Machida}, {Nakamura}, {Kudoh},
  {Akahori}, {Sofue}, \& {Matsumoto}}]{mac13}
{Machida} M., {Nakamura} K.~E., {Kudoh} T., {Akahori} T., {Sofue} Y.,
  {Matsumoto} R., 2013, \apj, 764, 81

\bibitem[{{Matsumoto} {et~al}\mbox{.}(1982){Matsumoto}, {Hayakawa}, {Koizumi},
  {Murakami}, {Uyama}, {Yamagami}, \& {Thomas}}]{mat82}
{Matsumoto} T., {Hayakawa} S., {Koizumi} H., {Murakami} H., {Uyama} K.,
  {Yamagami} T., {Thomas} J.~A., 1982, in American Institute of Physics
  Conference Series, Vol.~83, The Galactic Center, {Riegler} G.~R., {Blandford}
  R.~D., eds., pp. 48--52

\bibitem[{{McNally} \& {Pessah}(2014)}]{mp14}
{McNally} C.~P., {Pessah} M.~E., 2014, ArXiv e-prints

\bibitem[{{Miyamoto} \& {Nagai}(1975)}]{mn75}
{Miyamoto} M., {Nagai} R., 1975, \pasj, 27, 533

\bibitem[{{Molinari} {et~al}\mbox{.}(2011){Molinari}, {Bally},
  {Noriega-Crespo}, {Compi{\`e}gne}, {Bernard}, {Paradis}, {Martin}, {Testi},
  {Barlow}, {Moore}, {Plume}, {Swinyard}, {Zavagno}, {Calzoletti}, {Di
  Giorgio}, {Elia}, {Faustini}, {Natoli}, {Pestalozzi}, {Pezzuto},
  {Piacentini}, {Polenta}, {Polychroni}, {Schisano}, {Traficante}, {Veneziani},
  {Battersby}, {Burton}, {Carey}, {Fukui}, {Li}, {Lord}, {Morgan}, {Motte},
  {Schuller}, {Stringfellow}, {Tan}, {Thompson}, {Ward-Thompson}, {White}, \&
  {Umana}}]{mol11}
{Molinari} S. {et~al.}, 2011, \apjl, 735, L33

\bibitem[{{Morris} {et~al}\mbox{.}(1992){Morris}, {Davidson}, {Werner},
  {Dotson}, {Figer}, {Hildebrand}, {Novak}, \& {Platt}}]{mor92}
{Morris} M., {Davidson} J.~A., {Werner} M., {Dotson} J., {Figer} D.~F.,
  {Hildebrand} R., {Novak} G., {Platt} S., 1992, \apjl, 399, L63

\bibitem[{{Morris} \& {Serabyn}(1996)}]{ms96}
{Morris} M., {Serabyn} E., 1996, \araa, 34, 645

\bibitem[{{Morris}, {Uchida} \& {Do}(2006){Morris}, {Uchida}, \& {Do}}]{mor06}
{Morris} M., {Uchida} K., {Do} T., 2006, \nat, 440, 308

\bibitem[{{Morris}(2014)}]{mor14}
{Morris} M.~R., 2014, ArXiv e-prints

\bibitem[{{Nagayama} {et~al}\mbox{.}(2007){Nagayama}, {Omodaka}, {Handa},
  {Iahak}, {Sawada}, {Miyaji}, \& {Koyama}}]{nag07}
{Nagayama} T., {Omodaka} T., {Handa} T., {Iahak} H.~B.~H., {Sawada} T.,
  {Miyaji} T., {Koyama} Y., 2007, \pasj, 59, 869

\bibitem[{{Nishikori}, {Machida} \& {Matsumoto}(2006){Nishikori}, {Machida}, \&
  {Matsumoto}}]{nis06}
{Nishikori} H., {Machida} M., {Matsumoto} R., 2006, \apj, 641, 862

\bibitem[{{Nishiyama} {et~al}\mbox{.}(2010){Nishiyama}, {Hatano}, {Tamura},
  {Matsunaga}, {Yoshikawa}, {Suenaga}, {Hough}, {Sugitani}, {Nagayama}, {Kato},
  \& {Nagata}}]{nis10}
{Nishiyama} S. {et~al.}, 2010, \apjl, 722, L23

\bibitem[{{Oka} {et~al}\mbox{.}(2005){Oka}, {Geballe}, {Goto}, {Usuda}, \&
  {McCall}}]{oka05}
{Oka} T., {Geballe} T.~R., {Goto} M., {Usuda} T., {McCall} B.~J., 2005, \apj,
  632, 882

\bibitem[{{Oka} {et~al}\mbox{.}(1998){Oka}, {Hasegawa}, {Sato}, {Tsuboi}, \&
  {Miyazaki}}]{oka98}
{Oka} T., {Hasegawa} T., {Sato} F., {Tsuboi} M., {Miyazaki} A., 1998, \apjs,
  118, 455

\bibitem[{{Okuda} {et~al}\mbox{.}(1977){Okuda}, {Maihara}, {Oda}, \&
  {Sugiyama}}]{oku77}
{Okuda} H., {Maihara} T., {Oda} N., {Sugiyama} T., 1977, \nat, 265, 515

\bibitem[{{Parker}(1966)}]{pak66}
{Parker} E.~N., 1966, \apj, 145, 811

\bibitem[{{Peters}(1975)}]{pet75}
{Peters}, III W.~L., 1975, \apj, 195, 617

\bibitem[{{Riquelme} {et~al}\mbox{.}(2010){Riquelme}, {Bronfman},
  {Mauersberger}, {May}, \& {Wilson}}]{req10}
{Riquelme} D., {Bronfman} L., {Mauersberger} R., {May} J., {Wilson} T.~L.,
  2010, \aap, 523, A45

\bibitem[{{Rodriguez-Fernandez} \& {Combes}(2008)}]{rod08}
{Rodriguez-Fernandez} N.~J., {Combes} F., 2008, \aap, 489, 115

\bibitem[{{Rougoor} \& {Oort}(1960)}]{ro60}
{Rougoor} G.~W., {Oort} J.~H., 1960, Proceedings of the National Academy of
  Science, 46, 1

\bibitem[{{Sakamoto} {et~al}\mbox{.}(2006){Sakamoto}, {Ho}, {Iono}, {Keto},
  {Mao}, {Matsushita}, {Peck}, {Wiedner}, {Wilner}, \& {Zhao}}]{sak06}
{Sakamoto} K. {et~al.}, 2006, \apj, 636, 685

\bibitem[{{Sano}, {Inutsuka} \& {Miyama}(1999){Sano}, {Inutsuka}, \&
  {Miyama}}]{san99}
{Sano} T., {Inutsuka} S., {Miyama} S.~M., 1999, in Astrophysics and Space
  Science Library, Vol. 240, Numerical Astrophysics, {Miyama} S.~M., {Tomisaka}
  K., {Hanawa} T., eds., Boston, MA: Kluwer, p. 383

\bibitem[{{Sano} \& {Inutsuka}(2001)}]{sano01}
{Sano} T., {Inutsuka} S.-i., 2001, \apjl, 561, L179

\bibitem[{{Sawada} {et~al}\mbox{.}(2004){Sawada}, {Hasegawa}, {Handa}, \&
  {Cohen}}]{saw04}
{Sawada} T., {Hasegawa} T., {Handa} T., {Cohen} R.~J., 2004, \mnras, 349, 1167

\bibitem[{{Scoville}(1972)}]{sco72}
{Scoville} N.~Z., 1972, \apjl, 175, L127

\bibitem[{{Shakura} \& {Sunyaev}(1973)}]{ss73}
{Shakura} N.~I., {Sunyaev} R.~A., 1973, \aap, 24, 337

\bibitem[{{Shibata} \& {Matsumoto}(1991)}]{sm91}
{Shibata} K., {Matsumoto} R., 1991, \nat, 353, 633

\bibitem[{{Sofue}(2007)}]{sof07}
{Sofue} Y., 2007, \pasj, 59, 189

\bibitem[{{Sofue}(2013)}]{sof13}
{Sofue} Y., 2013, \pasj, 65, 118

\bibitem[{{Stone} \& {Norman}(1992)}]{sn92}
{Stone} J.~M., {Norman} M.~L., 1992, \apjs, 80, 791

\bibitem[{{Suzuki} \& {Inutsuka}(2005)}]{si05}
{Suzuki} T.~K., {Inutsuka} S.-i., 2005, \apjl, 632, L49

\bibitem[{{Suzuki} \& {Inutsuka}(2006)}]{si06}
{Suzuki} T.~K., {Inutsuka} S.-i., 2006, Journal of Geophysical Research (Space
  Physics), 111, 6101

\bibitem[{{Suzuki} \& {Inutsuka}(2009)}]{si09}
{Suzuki} T.~K., {Inutsuka} S.-i., 2009, \apjl, 691, L49

\bibitem[{{Suzuki} \& {Inutsuka}(2014)}]{si14}
{Suzuki} T.~K., {Inutsuka} S.-i., 2014, \apj, 784, 121

\bibitem[{{Suzuki}, {Muto} \& {Inutsuka}(2010){Suzuki}, {Muto}, \&
  {Inutsuka}}]{suz10}
{Suzuki} T.~K., {Muto} T., {Inutsuka} S.-i., 2010, \apj, 718, 1289

\bibitem[{{Takahashi} {et~al}\mbox{.}(2009){Takahashi}, {Nozawa}, {Matsumoto},
  {Machida}, {Fukui}, {Kudo}, {Torii}, {Yamamoto}, \& {Fujishita}}]{tak09}
{Takahashi} K. {et~al.}, 2009, \pasj, 61, 957

\bibitem[{{Takeuchi} {et~al}\mbox{.}(2010){Takeuchi}, {Yamamoto}, {Torii},
  {Kudo}, {Hayakawa}, {Kawamura}, {Mizuno}, {Onishi}, {Mizuno}, {Ogawa}, \&
  {Fukui}}]{tak10}
{Takeuchi} T. {et~al.}, 2010, \pasj, 62, 557

\bibitem[{{Torii} {et~al}\mbox{.}(2014{\natexlab{a}}){Torii}, {Enokiya},
  {Fukui}, {Yamamoto}, {Kawamura}, {Mizuno}, {Onishi}, \& {Ogawa}}]{tor14a}
{Torii} K., {Enokiya} R., {Fukui} Y., {Yamamoto} H., {Kawamura} A., {Mizuno}
  N., {Onishi} T., {Ogawa} H., 2014{\natexlab{a}}, in IAU Symposium, Vol. 303,
  IAU Symposium, {Sjouwerman} L.~O., {Lang} C.~C., {Ott} J., eds., pp. 106--108

\bibitem[{{Torii} {et~al}\mbox{.}(2014{\natexlab{b}}){Torii}, {Enokiya},
  {Morris}, {Hasegawa}, {Kudo}, \& {Fukui}}]{tor14b}
{Torii} K., {Enokiya} R., {Morris} M.~R., {Hasegawa} K., {Kudo} N., {Fukui} Y.,
  2014{\natexlab{b}}, \apjs, 213, 8

\bibitem[{{Torii} {et~al}\mbox{.}(2010{\natexlab{a}}){Torii}, {Kudo},
  {Fujishita}, {Kawase}, {Okuda}, {Yamamoto}, {Kawamura}, {Mizuno}, {Onishi},
  {Machida}, {Takahashi}, {Nozawa}, {Matsumoto}, {Ott}, {Tanaka}, {Yamaguchi},
  {Ezawa}, {Stutzki}, {Bertoldi}, {Koo}, {Bronfman}, {Burton}, {Benz}, {Ogawa},
  \& {Fukui}}]{tor10a}
{Torii} K. {et~al.}, 2010{\natexlab{a}}, \pasj, 62, 675

\bibitem[{{Torii} {et~al}\mbox{.}(2010{\natexlab{b}}){Torii}, {Kudo},
  {Fujishita}, {Kawase}, {Yamamoto}, {Kawamura}, {Mizuno}, {Onishi}, {Mizuno},
  {Machida}, {Takahashi}, {Nozawa}, {Matsumoto}, \& {Fukui}}]{tor10b}
{Torii} K. {et~al.}, 2010{\natexlab{b}}, \pasj, 62, 1307

\bibitem[{{Tsuboi}, {Handa} \& {Ukita}(1999){Tsuboi}, {Handa}, \&
  {Ukita}}]{tsu99}
{Tsuboi} M., {Handa} T., {Ukita} N., 1999, \apjs, 120, 1

\bibitem[{{Tsuboi} {et~al}\mbox{.}(1986){Tsuboi}, {Inoue}, {Handa}, {Tabara},
  {Kato}, {Sofue}, \& {Kaifu}}]{tsu86}
{Tsuboi} M., {Inoue} M., {Handa} T., {Tabara} H., {Kato} T., {Sofue} Y.,
  {Kaifu} N., 1986, \aj, 92, 818

\bibitem[{{Velikhov}(1959)}]{vel59}
{Velikhov} E.~P., 1959, Zh. Eksp. Teor. Fiz., 36, 1398

\bibitem[{{Yusef-Zadeh}, {Morris} \& {Chance}(1984){Yusef-Zadeh}, {Morris}, \&
  {Chance}}]{yus84}
{Yusef-Zadeh} F., {Morris} M., {Chance} D., 1984, \nat, 310, 557

\end{thebibliography}
\end{document}